\documentclass[sigconf]{acmart}
\setcopyright{none} 

\acmConference{}{}{}
\renewcommand\footnotetextcopyrightpermission[1]{}

\acmYear{2021}
\author{Florian Sieck}
\email{florian.sieck@uni-luebeck.de}
\orcid{ORCID}
\affiliation{%
  \institution{University of L{\"u}beck}
  \department{Institute for IT Security}
  \country{Germany}
}

\author{Sebastian Berndt}
\email{s.berndt@uni-luebeck.de}
\orcid{0000-0003-4177-8081}
\affiliation{%
  \institution{University of L{\"u}beck}
  \department{Institute for IT Security}
  \country{Germany}
}

\author{Jan Wichelmann}
\email{j.wichelmann@uni-luebeck.de}
\orcid{ORCID}
\affiliation{%
  \institution{University of L{\"u}beck}
  \department{Institute for IT Security}
  \country{Germany}
}

\author{Thomas Eisenbarth}
\email{thomas.eisenbarth@uni-luebeck.de}
\affiliation{%
  \institution{University of L{\"u}beck}
  \department{Institute for IT Security}
  \country{Germany}
}

\usepackage{tikz}
\usetikzlibrary{shapes, arrows, positioning}
\usepackage{amsmath}

\usepackage{amsthm}
\usepackage[nolist]{acronym}
\usepackage{listings}
\usepackage{mdframed}
\usepackage{color}
\newtheorem*{assumption}{Assumption}
\newtheorem{theorem}{Theorem}
\newtheorem{lemma}[theorem]{Lemma}
\usepackage{graphicx}
\usepackage{pdfpages}
\usepackage{url}
\usepackage{xspace}
\usepackage{enumitem}

\usepackage{stfloats}
\usepackage{bm}

\usepackage{cryptocode}
\usepackage{booktabs}

\usepackage{threeparttable}

\usepackage[htt]{hyphenat}

\newcommand{\enc}[1]{\langle #1\rangle}
\newcommand{\sk}{\textit{sk}}
\newcommand{\pr}{\textit{pr}}
\DeclareMathOperator{\prob}{\mathrm{Pr}}
\newcommand{\rhs}{\operatorname{rhs}}

\newcommand{\Vars}{\mathsf{Vars}}

\newcommand{\obs}{\operatorname{\text{\textsc{obs}}}}
\newcommand{\block}{\operatorname{block}}
\newcommand{\parts}{\operatorname{\text{\textsc{part}}}}

\newcommand{\FlushReload}{Flush+\allowbreak Reload\xspace}
\newcommand{\PrimeProbe}{Prime+\allowbreak Probe\xspace}

\newcommand{\basesf}{base64\xspace}

\newcommand{\openssl}{OpenSSL\xspace}
\newcommand{\wolfssl}{wolfSSL\xspace}
\newcommand{\botan}{Botan\xspace}
\newcommand{\nss}{NSS\xspace}
\newcommand{\gnunettle}{GNU Nettle\xspace}
\newcommand{\rustsgx}{Rust\allowbreak SGX\xspace}
\newcommand{\mbedtls}{mbed\allowbreak TLS\xspace}
\newcommand{\boringssl}{BoringSSL\xspace}
\newcommand{\talos}{TaLoS\xspace}

\makeatletter
\newcommand*\mywidetilde[1]{\mathpalette\wthelper{#1}}
\newcommand*\wthelper[2]{%
        \hbox{\dimen@\accentfontxheight#1%
                \accentfontxheight#11.3\dimen@
                $\m@th#1\widetilde{#2}$%
                \accentfontxheight#1\dimen@
        }%
}

\newcommand*\accentfontxheight[1]{%
        \fontdimen5\ifx#1\displaystyle
                \textfont
        \else\ifx#1\textstyle
                \textfont
        \else\ifx#1\scriptstyle
                \scriptfont
        \else
                \scriptscriptfont
        \fi\fi\fi3
}
\makeatother

\begin{document}

\date{}

\title[Util::Lookup: Exploiting key decoding in cryptographic libraries]{Util::Lookup: Exploiting key decoding in cryptographic libraries}

\begin{acronym}
  \acro{ASN.1}{Abstract Syntax Notation One}
  \acro{BER}{Basic Encoding Rules}
  \acro{DER}{Distinguished Encoding Rules}
  \acro{PEM}{Privacy-enhanced Electronic Mail}
  \acro{CRT}{Chinese Remainder Theorem}
  \acro{SGX}{Software Guard Extensions}
  \acro{LUT}{lookup table}
  \acro{MI}{Mutual Information}
  \acro{MSR}{Model-specific register}
  \acro{PSW}{Platform Software}
  \acro{SDK}{Software Development Kit}
  \acro{LVI}{Load Value Injection}
  \acro{LLC}{last level cache}
  \acro{EPC}{Enclave Page Cache}
  \acro{PTE}{page table entry}
  \acro{EPCM}{Enclave Page Cache Map}
  \acro{AEP}{Asynchronous Exit handler Pointer}
  \acro{AEX}{Asynchronous Enclave Exit}
  \acro{L1TF}{L1 Terminal Fault}
  \acro{SMT}{Simultaneous Multithreading}
  \acro{OS}{operating system}
\end{acronym}

\definecolor{listinggreen}{rgb}{0,0.6,0}
\definecolor{verylightgray}{rgb}{0.97,0.97,0.97}
\definecolor{customdarkgray}{rgb}{0.35, 0.35, 0.35}
\definecolor{violett}{rgb}{0.3, 0.0, 0.4}
\lstset{
	basicstyle=\small\ttfamily,
	morecomment=[l][\color{listinggreen}]{\#},
	escapeinside={(*@}{@*)}, 
}

\begin{abstract}
Implementations of cryptographic libraries have been scrutinized for secret-dependent execution behavior exploitable by microarchitectural side-channel attacks. To prevent unintended leakages, most libraries moved to constant-time implementations of cryptographic primitives. There have also been efforts to certify libraries for use in sensitive areas, like Microsoft CNG and Botan, with specific attention to leakage behavior.

In this work, we show that a common oversight in these libraries is the existence of \emph{utility functions}, which handle and thus possibly leak confidential information. 
We analyze the exploitability of base64 decoding functions across several widely used cryptographic libraries. Base64 decoding is used when loading keys stored in PEM format. We show that these functions by themselves leak sufficient information even if libraries are executed in trusted execution environments. In fact, we show that recent countermeasures to transient execution attacks such as LVI \emph{ease} the exploitability of the observed faint leakages, allowing us to robustly infer sufficient information about RSA private keys \emph{with a single trace}. We present a complete attack, including a broad library analysis, a high-resolution last level cache attack on SGX enclaves, and a fully parallelized implementation of the extend-and-prune approach that allows a complete key recovery at medium costs.
\end{abstract}

    
	\maketitle
	
\section{Introduction}

Due to the widespread adoption of cloud-computing and virtual machines, architectural and microarchitectural attacks exploiting shared resources have become a major concern for security-critical applications~\cite{zhang2012cross,liu2015last,inci2016cache}. 
Within the last decade, these attacks have seen great advances~\cite{percival2005cache, irazoqui2015s, yarom2014flush, moghimi2020copycat}, culminating in jeopardizing the security of trusted execution environments, cloud computing, and finally revealing transient execution bugs in all modern processors~\cite{lipp2018meltdown, kocher2019spectre, van2018foreshadow, van2020lvi}.

A popular target of these attacks are cryptographic implementations, as they contain critical information that is both compact and used for extensive and often highly optimized computations. 
As a result, cryptographic implementations have been analyzed for exploitable code behavior that leaks information in great detail. This task has been significantly eased by the development of more automated analysis techniques, which are offered by tools like \emph{CacheAudit}~\cite{doychev2015cacheaudit}, \emph{DATA}~\cite{WeiserDATA} or \emph{Microwalk}~\cite{wichelmann2018microwalk}. In fact, recent studies have identified remaining exploitable code sections in cryptographic implementations, with increasingly small leakages \cite{aranha2020ladderleak,moghimi2020copycat,hassan2020deja,weiser2020big} and have resulted in a long stream of CVEs for these libraries. These remaining leakages are getting smaller and more difficult to find due to the vast effort that went into the analysis. Some libraries have even been certified for secure implementation and design: \emph{Microsoft Cryptography API: Next Generation} (CNG) is periodically FIPS-validated~\cite{cngfips}; \emph{Botan} has been extensively audited to be resistant against common side-channel attacks, and is thus authorized for use in sensitive applications~\cite{BSIBotan}.

At the same time, Intel has released numerous mitigations for transient execution bugs which significantly altered the efficacy of microarchitectural attacks, in particular in highly protected environments such as Intel Software Guard Extensions (SGX). Countermeasures to Foreshadow \cite{van2018foreshadow}, also known as L1 Terminal Fault, now prevent L1 Cache attacks on SGX~ \cite{deepdivel1tf}, while attacks exploiting \ac{SMT} should be prevented by enclave developers ensuring operation on private cores.

\subsection{Our Contribution}

In this work, we show that despite the increasingly rigorous analysis of cryptographic libraries, microarchitectural attacks are still a threat. Our analysis finds that remaining issues are not in the  cryptographic routines that have been extensively analyzed by other publications. Instead, we investigate \emph{utility functions}, which are a vital part of cryptographic libraries, but have thus far been ignored in most studies.
Clearly, these functions compute on sensitive data and are thus potential subject to information leakage. 

In particular, we focus on functionality that decodes secret keys from a format suitable for storage and converts them into binary data used at runtime. 
We show how such a functionality can be exploited via a faint \ac{LLC} leakage.

More concretely, we investigate RSA keys that are stored in the popular \acs{PEM} format, which uses \basesf to encode binary data in printable characters.
The decoded information is later processed in the constant-time cryptographic implementations of the library.

Usually, the leakage of the decoding process is quite faint and extremely difficult to exploit: Subsequent table accesses during decoding are only few instructions apart and will be executed out-of-order, resulting in extremely high noise, while featuring only minimal leakage to begin with.
In fact, even a powerful attack that combines three different microarchitectural attack techniques, the page fault side channel~\cite{xu2015controlled}, the single-stepping of SGX-step~\cite{vanbulck2017SGXStep}, and a last-level cache attack for spatial resolution of cache accesses~\cite{liu2015last} has problems to distinguish two close cache hits in the \ac{LLC} due to the noise. 
Thus, other studies which already noticed such leakages in key decoding~\cite{WeiserDATA} ignored these findings, likely because they were not deemed exploitable, and the vulnerable code remained unfixed.

Many of the recent microcode updates render side-channel attacks on SGX more difficult, in particular by flushing the L1 data cache and reflecting the \ac{SMT} state in the attestation.
However, it turns out that the recent mitigation against \ac{LVI} and the resulting serialization of memory accesses in SGX enclaves enables us to sample the decoding with significantly reduced noise and to obtain \emph{almost error-free leakage}, even from a \emph{single} key loading event. 
To practically exploit the observed leakages, we make use of state-of-the-art cryptanalytic methods. 

In summary our contributions are:

\begin{itemize}

    \item Fine-grained leakage analysis of the \basesf decoding functions for several common crypto libraries, including the certified \botan library.
    \item A sophisticated microarchitectural attack that manages to extract the observed leakages from SGX enclaves with a single observation, in spite of and because of the microcode and  countermeasures introduced by Intel in response to transient execution attacks.
    \item An optimized RSA key recovery method including a highly scalable implementation that, given medium resources, allows to reconstruct the key from very weak leakages of 5 of the 6 redundant key parameters commonly used for RSA key storage, as well as a thorough performance analysis.
\end{itemize}

We plan to publish the complete code of our attack.

\subsection{Responsible Disclosure}
We have informed the maintainers of all studied libraries about our findings in December 2020.
\begin{itemize}
    \item Botan: CVE-2021-24115, fixed with version 2.17.3
    \item GNU Nettle: No response, not yet fixed
    \item mbedTLS: CVE-2021-24119, fixed with version 2.26.0
    \item MS CryptoAPI: Declared as not urgent
    \item NSS: Not yet fixed
    \item OpenSSL: No response, not yet fixed
    \item RustSGX: CVE-2021-24117, pending fix
    \item wolfSSL: CVE-2021-24116, fixed with version 4.6.0
\end{itemize}

\section{Background}

\subsection{Microarchitectural Attacks}
A common approach in microarchitectural attacks is the exploitation of contention in microarchitectural buffers within the CPU, which are intended to improve performance. The usually unprivileged attacker manipulates the content of a buffer or cache to provoke abnormal behavior, which can be measured as side-channel information like timing behavior, and leak  secrets partially.

\subsubsection{Cache attacks} A frequent target are CPU caches, from the core-specific L1 caches \cite{percival2005cache, bernstein2005cache}, to the unified, inclusive and shared L3 caches \cite{liu2015last, irazoqui2015s}. Many techniques evolved, allowing the adversary to measure timing differences regarding data or instruction accesses \cite{yarom2014flush, osvik2006cache, gruss2016flush, briongos2020reload+}, in order to determine whether the victim code accessed a certain cache line.

Intel's processors have a set-associative cache layout, where each 64-byte cache line maps into a specific  \emph{cache set}. Each cache set has a limited number of \emph{ways}, which is the number of cache lines it can contain at any time. The cache sets of the L3 cache are divided into \emph{slices}, where the number of slices usually corresponds to the number of logical cores. The mapping of physical address to slices is computed by an undocumented hash function.

\PrimeProbe \cite{osvik2006cache} determines a victim's cache access by first priming a complete cache set with attacker values, the so-called eviction set, then waiting on the victim's code execution and finally probing the complete cache set with the same attacker values. If the probing access time is below a threshold, all attacker values were served from the cache and the victim did not access the data of interest. Otherwise, if the access time is above that threshold, the victim accessed data that was mapped to the same cache set and thus evicted some of the attacker's values. Unlike \FlushReload~\cite{yarom2014flush}, this procedure does not require any shared memory, but it is less precise and prone to noise, since a complete set is probed instead of a single cache line. When attacking the sliced L3 cache, it is advisable to use one eviction set per slice, to reduce noise caused by the remaining system and its processes. \PrimeProbe first requires to construct eviction set(s), meaning finding $c_{size}$ addresses which map to the cache set that is to be primed, where $c_{size}$ specifies the size of a cache set. Depending on the attacker model, this can be done using virtual to physical address translation, or through huge pages usually giving the attacker control over the cache set index bits \cite{liu2015last, vila2019theory}. Constructing eviction sets per slice does not require the knowledge of the address to slice mapping. Instead they can be derived incrementally from the cache set's eviction set~\cite{liu2015last}.

\subsubsection{Vulnerability Detection} Finding side-channel vulnerabilities in programs by manually inspecting high level and assembly code is a cumbersome task and will reveal only a small portion of vulnerabilities, and has to be repeated each time the code was changed.

Automated vulnerability finding can assist in this endeavor. One approach to this end is to leverage dynamic binary instrumentation and analysis combined with input fuzzing, as done by \textit{DATA}~\cite{WeiserDATA} and \textit{Microwalk}~\cite{wichelmann2018microwalk}. These frameworks find non-constant-time behavior by instrumenting a piece of code under test and executing it multiple times with different inputs (secrets), while recording the execution traces. Deviations between traces suggest secret dependent behavior. The tools quantify the leakages by calculating the mutual information between input and observed traces.

There also are different approaches like e.\,g.~CacheAudit \cite{doychev2015cacheaudit}, which applies static analysis to find cache side channels.

\subsection{Intel SGX}

Running security relevant software or algorithms processing confidential data in untrusted environments has become quite  common. Intel \ac{SGX} aims to provide a hardware root of trust, enabling users to run software in isolated environments, called \emph{enclaves}, which can perform confidential computations in the presence of an untrusted \ac{OS} without leaking secret information \cite{hoekstra2013using, mckeen2013innovative}. To allow the application developer to verify the integrity and security of their application, Intel SGX supports two remote attestation schemes \cite{anati2013innovative, scarlata2018supporting}.

\subsubsection{Memory management} 
Intel \ac{SGX} leaves the memory management to the \ac{OS}, which is responsible for allocating memory and mapping physical to virtual  memory addresses. This allows untrusted software to tamper with \ac{PTE} meta information like the \ac{PTE} accessed bit \cite{vanbulck2017SGXStep}. However, Intel SGX specifically guarantees integrity and confidentiality of the data in RAM. All data in RAM is protected by memory encryption \cite{gueron2016memory} and kept in the \ac{EPC}, which is inaccessible from outside \ac{SGX}. 
To counteract manipulations of the 
address translation, \ac{SGX} keeps track of all \ac{EPC} memory pages including their expected virtual address in the \ac{EPCM}~\cite{costan2016intel}.

\subsubsection{Context switches} Programs running in Intel \ac{SGX} are subject to context switches as any other process. Since enclaves are isolated from the remaining system, context switch require dedicated instructions provided by \ac{SGX}~\cite{mckeen2013innovative, costan2016intel}. After an enclave was created with the \texttt{ECREATE} instruction, it can be entered with \texttt{EENTER}. In case the processor is interrupted while in enclave mode, \ac{SGX} ensures that \texttt{\ac{AEX}} is executed, storing the execution state to a secure area and cleaning up the registers. Additionally, the instruction pointer is set to the \ac{AEP}, causing the system's interrupt handler to return to \ac{AEP} when it finishes. Finally, \texttt{ERESUME} can be called from the asynchronous exit handler to resume execution of the enclave.

\subsubsection{Attacks on Intel \ac{SGX}}
Trusted execution environments like Intel \ac{SGX}  feature an attacker model which assumes an untrusted \ac{OS}, and thus enables adversaries to tamper with \emph{all} system resources to extract information from isolated enclaves.

Amplifying side-channels with control over system events and resources, such as page faults or interrupts, and using this to reduce noise, is called a controlled-channel attack~\cite{xu2015controlled}.

\textit{SGX-Step}~\cite{vanbulck2017SGXStep} introduced a framework for controlled-channel attacks on \ac{SGX}, which was used in many subsequent attacks~\cite{van2018nemesis, moghimi2020copycat, van2020lvi, aldaya2020one}. It enables the attacker to single step enclaves and to manipulate page table entries in order to get insight into the control flow.

Transient execution attacks on \ac{SGX}~\cite{van2018foreshadow, van2020lvi, SGXpectre18, schwarz2019zombieload, ragab2021crosstalk} have forced Intel to publish microcode and software mitigations.  One countermeasure pushed via microcode updates is to flush microarchitectural buffers such as the L1 data cache upon enclave exit \cite{deepdivel1tf}. In addition, compilers now insert fences in enclave code to prevent Spectre-like attacks. Furthermore, disabling simultaneous multithreading is recommended when executing enclaves.
In addition to transient execution attacks and controlled-channel attacks, other vulnerabilities were found targeting e.\,g.~the cache \cite{moghimi2017cachezoom, dall2018cachequote} or the branch history \cite{lee2017inferring}.

\subsection{RSA Key recovery}
Recovering the complete RSA key from partial information has been studied in
numerous settings.
In theory, it is sufficient to store only one of the primes $p$ or $q$ as
private key, but this is very inefficient.
To speed up the decryption of messages via the \ac{CRT}, \emph{all} of the values
$(p,q,d,d_{p},d_{q},q_{p}^{-1})$ are stored, where $d_{p}:=d \pmod{p-1}$, $d_{q}
:= d \pmod{q-1}$, and $(q_{p}^{-1}\cdot q ) \pmod{p} = 1$.
Note that the knowledge of any single of these variables is sufficient to
reconstruct all other variables, given the public key
$(N,e)$~\cite{heninger2009reconstructing}.

There are roughly two kinds of partial information that are obtained by
side-channel attacks: \emph{consecutive} information and \emph{non-consecutive}
information.
In the consecutive case, the attacker obtains a few number of consecutive blocks of information
about some of the variables, e.\,g.~the $\enc{p}/2$ most significant bits of
$p$, where $\enc{p}$ is the \emph{encoding length} of $p$, i.\,e.~$\enc{p} :=
\lceil \log_{2}(p+1) \rceil$ or the $\enc{p}/4$ most significant bits of $p$ and the $\enc{p}/4$ least significant bits of~$p$. 
This continuity gives a high amount of structured information, which allows
an attacker to mount attacks based on \emph{lattices}.
For the many applications of this technique to reconstruct parts of the private
key, we refer to the  surveys~\cite{boneh1999twenty,DBLP:series/isc/May10,
  DBLP:conf/ctrsa/TakayasuK17}.
This technique was also used in a recent work to recover RSA keys~\cite{moghimi2020medusa}, where the authors were able to obtain partial leakages of $q$ on consecutive positions and could use this information to derive $q$ completely.  

In the non-consecutive case, the information is widely spread over the
variables.

A widely used algorithm for this case was presented by Heninger and Shacham and
subsequently generalized by Henecka et al.~and Paterson et
al.~\cite{DBLP:conf/asiacrypt/PatersonPS12,DBLP:conf/crypto/HeneckaMM10,heninger2009reconstructing}.
In a naive fashion, one could try to construct a search tree that aims to test
all possibilities for the $6$ different unknown variables, which gives a solution space of
$2^{6n}$.
Whenever a candidate is encountered that does not fit to the partial known
information, we can prune this candidate.
The main idea of Heninger and Shacham is to use the different dependencies
between the variables to set up an equation system containing $4$ equations and
$5$ variables, which drastically reduces the solution space to $2^{n}$.
Given sufficient information from side-channel attacks can then be used to further reduce this space. 
This approach was used for many attacks, e.\,g.~\cite{DBLP:conf/ches/BernsteinBGBHLV17,yarom2017cachebleed,brasser2017software,gras2018translation}.
In all variations of the algorithm, the partial information contains information
on a bit-wise level, while our attack works on information about blocks of bits. 
We thus adapt the algorithm of Heninger and Shacham to this setting in
Section~\ref{sec:key_recovery}. 

\section{Exploiting Key Decoding}
\label{sec:base64_decoding}

In this section we analyze possible leakages in the key decoding routines of various cryptographic libraries. First, we describe the \ac{PEM} format, which is used for storing and exchanging cryptographic material, and is supported by many common cryptographic implementations. We use the \emph{Microwalk}~\cite{wichelmann2018microwalk} framework to conduct a broad analysis of several popular libraries, including {\openssl}~\cite{opensslgit}, {\wolfssl}~\cite{wolfsslgit}, {\nss}~\cite{nssgit}, and {\botan}~\cite{botangit} in order to find and assess possible leakages in key decoding. Microsoft CNG itself does not offer native key decoding, and offloads this onto the user; however, its largely deprecated predecessor \emph{Microsoft Crypto API}~\cite{cryptoapi} is still included in recent Windows versions, and supports loading and storing \ac{PEM} formatted keys. WolfSSL, {\rustsgx}~\cite{rustsgxgit} and \mbedtls~\cite{mbedtlsgit} offer native SGX support, and with {\talos}~\cite{talosgit} there also is an SGX mode for \openssl. Finally, we analyze \emph{Microwalk}'s findings and show that \ac{LUT}-based \basesf decoding poses a significant and widespread source of leakage, which we exploit to infer the entire private key in Sections \ref{sec:cache_attack_in_intel_sgx} and \ref{sec:key_recovery}.

\subsection{Storing Cryptographic Material}
Storage formats for cryptographic data face several challenges: The format should be standardized, such that it can be exchanged between different implementations without compatibility issues. Then, fingerprints of keys and certificates should be unambiguous, i.\,e., there shouldn't be two equivalent representations of the same cryptographic entity. Finally, while not a hard requirement, the format should be easily usable in practice, to allow transferring cryptographic data without worrying about encoding issues.

\subsubsection{PEM Format}
To accomplish this, RSA private keys are commonly stored in PKCS \#8 format~\cite{pkcs8}, which is specified in \ac{ASN.1} interface description language~\cite{asn1} and uses the \ac{DER} encoding to generate a unique binary representation for cryptographic data. This encoding is defined in such a way that it is ensured that the same key material always yields the same binary data. Listing \ref{lst:rsa_key_der} in the appendix shows an example 1024-bit RSA private key in \ac{ASN.1} format, encoded with \ac{DER}. Data encoded with \ac{DER} can be encrypted using a symmetric algorithm and wrapped into another \ac{DER} layer, to protect it in case the key file gets stolen; however, for server deployments, the same applies for the used passphrase, which usually is stored next to the encrypted key file, limiting the security benefit. For this reason, unencrypted key files are still prevalent.

Finally, in order to allow easy handling and transmission over non-binary channels, the binary \ac{DER} data is \basesf encoded and complemented with start and end markers, which denote the semantics of the {\basesf}-encoded payload, and allow implementations to easily determine the correct decoding technique. These markers also allow to store multiple entities in one file, e.\,g.~certificate files, containing certificates of an entire chain. These files are usually referred to as \ac{PEM} format.

\subsubsection{Encoded RSA Private Keys: }
An RSA private key typically consists of the public parameters $N$ and $e$, as well as the private parameter $d$. These values are sufficient for decrypting and signing messages. For better performance, many implementations utilize the \ac{CRT}, which additionally requires the primes $p$ and $q$, and three parameters $d_p = d\bmod (p-1)$, $d_q = d\bmod (q-1)$ and $q_{inv} = q^{-1} \bmod p$. 

\subsection{Finding Leakages}

\subsubsection{Leakage Detection}
In order to avoid time-consuming and error-prone manual analysis, we utilized the \emph{Microwalk}~\cite{wichelmann2018microwalk} framework to automatically analyze the key decoding of several major cryptographic libraries, and infer possibly interesting leakages. This approach has the advantage that we can focus on the code sections which actually do behave differently depending on the secret input (and thus may leak), and it also finds very subtle leakages often missed when doing manual analysis, but exploitable nonetheless.

Since \emph{Microwalk} relies on dynamic instrumentation, we randomly generated a set of 4,096 private key PEM files with slightly varying parameter sizes, and traced the key decoding of each library. We then instructed the analysis module to compute the amount of leaked bits per memory accessing instruction. After \emph{Microwalk} had generated and analyzed the traces for each test case, we manually removed false positives like subtle variations in the memory allocator and reports relating to cryptographic operations, and sorted the results by their estimated severity.

For \openssl, the resulting leakage candidates were all related to decoding the private key. Functions prefixed with the string \lstinline{EVP} were assigned the highest possible leakage estimation: The \lstinline{EVP_DecodeUpdate}

method does an initial scan of the entire input string, in order to determine its length and remove invalid characters, and then passes it to the \lstinline{EVP_DecodeBlock} method, which performs a \ac{LUT}-based \basesf decoding of the input. Another notable leakage is the \lstinline{BN_bin2bn} function, which converts the decoded key parameters into big number objects: It loops over the currently processed parameter, and thus leaks its length. Detailed analysis results for \openssl are listed in Table~\ref{tab:microwalk-results} in the appendix. We continue with explaining and discussing these leakages in detail.

\subsection{Analysis of Key Decoding Techniques}

\subsubsection{Decoding of PEM Files}
When loading the private key, cryptographic libraries parse the \ac{PEM} file, decode the \basesf \ac{DER}, and convert the binary \ac{DER} representation into an internal format. For those libraries that employ a \acf{LUT}-based approach, we found that in \emph{each} analyzed library this process leaks key information for every \basesf character, and thus every parameter stored in the key file.

All libraries roughly follow the same high-level approach: First, they parse the start/end markers to locate the {\basesf}-encoded payload. Then, they decode each {\basesf}-character and reconstruct the underlying binary data, while skipping invalid characters like line breaks and spaces. Finally, the DER container is handed to the next decoder stage, which parses the DER blocks following the ASN.1 specification, and initializes a corresponding private key object.

\subsubsection{Leakages in \basesf Decoding}
\label{subsubsec:leakages_in_base64_dec}
In \basesf encoding, the binary data is divided into 6-bit chunks, interpreted as alphanumeric characters, the plus sign or the slash, making up 64 distinct characters, all from the ASCII character set.

For decoding, these characters are converted back into 6-bit chunks, where each group of four chunks corresponds to 3 bytes of binary data. While this conversion can be realized as a case decision, most implementations rely on \ac{LUT}s, where each ASCII character maps to the corresponding 6-bit chunk (or an invalid value). Since an ASCII-encoded character takes up 7 bits, the \ac{LUT}s need to have at least 128 entries. Listing \ref{lst:openssl_lut} shows the decoding table used by \openssl.
\begin{figure}[t]
\begin{mdframed}[backgroundcolor=verylightgray]
\begin{lstlisting}[basicstyle=\scriptsize\ttfamily]
(*@\color{violett}{0x00}@*) ffffffff ffffffff 
(*@\color{violett}{0x08}@*) ffe0f0ff fff1ffff # TAB LF CR
(*@\color{violett}{0x10}@*) ffffffff ffffffff 
(*@\color{violett}{0x18}@*) ffffffff ffffffff
(*@\color{violett}{0x20}@*) e0ffffff ffffffff # SPACE
(*@\color{violett}{0x28}@*) ffffff3e fff2ff3f # + - /
(*@\color{violett}{0x30}@*) 34353637 38393a3b # 0 1 2 3 4 5 6 7 
(*@\color{violett}{0x38}@*) 3c3dffff ff00ffff # 8 9 =
(*@\color{violett}{0x40}@*) ff000102 03040506 # A B C D E F G
(*@\color{violett}{0x48}@*) 0708090a 0b0c0d0e # H I J K L M N O
(*@\color{violett}{0x50}@*) 0f101112 13141516 # P Q R S T U V W
(*@\color{violett}{0x58}@*) 171819ff ffffffff # X Y Z
(*@\color{violett}{0x60}@*) ff1a1b1c 1d1e1f20 # a b c d e f g
(*@\color{violett}{0x68}@*) 21222324 25262728 # h i j k l m n o
(*@\color{violett}{0x70}@*) 292a2b2c 2d2e2f30 # p q r s t u v w
(*@\color{violett}{0x78}@*) 313233ff ffffffff # x y z
\end{lstlisting}
\end{mdframed}
\caption{\basesf-decoding \acl{LUT} as present in the \openssl binary. The comment column on the right lists the ASCII representations of valid code points (non-\texttt{0xFF} bytes).}
\label{lst:openssl_lut}
\end{figure}
Note that due to its length, the table takes up at least two 64-byte cache lines, which allows an attacker to infer a part of the table index through a cache attack, as we will show in Section~\ref{sec:cache_attack_in_intel_sgx}. While all analyzed libraries use a \ac{LUT}-based {\basesf}-decoding, the exact implementations vary in detail: For example, \openssl and \nss parse the \basesf string twice, to handle invalid or white space characters, and determine the length of the resulting decoded binary string. This allows the attacker to do multiple measurements per input, which reduces the measurement error.

Another difference between the libraries and even between different configurations of a single library is the alignment of the \basesf \ac{LUT}. If a 128-byte \ac{LUT} is aligned at a cache line boundary (64 bytes), it takes up exactly two cache lines. As depicted in Listing~\ref{lst:openssl_lut}, the \ac{LUT} entries are not evenly distributed: Considering only the \basesf character set, the first half has 12 entries, while the second half has 52, so observing an access to the first cache line yields more information than an access to the second one. However, if the \ac{LUT} is aligned at 32 bytes, the entries are split over three cache lines: The first one does not have any \basesf entry, the second has 38, and the third has 26.

To measure the average information that is leaked by a \ac{LUT} access when observed at cache line level, we compare the number of \basesf entries per cache line. Let random variable $B$ denote the 64 possible \basesf characters, where each character $b$ has the same probability: $\prob [B=b]=\frac{1}{64}$. Also, let random variable $C$ denote the cache lines which contain a part of the \ac{LUT}. The probability that we observe a certain cache line $c$ is thus $\prob [C=c]$, which equals the fraction of \basesf characters which map to this cache line. Finally, $\prob [B=b\,|\,C=c]$ denotes the probability of a certain \basesf character $b$ if we observed cache line $c$.

We can then compute the average information $I(B,C)=H(B)-H(B\,|\,C)$ leaked by observing a cache line, where $H$ denotes the Shannon-entropy. 

Table \ref{tab:crypt_libraries_overview} shows the investigated libraries and the expected  leakage for \basesf decoding.

Note that the amount of leaked information depends on the structure and the alignment of the \ac{LUT}: If the \ac{LUT} takes up two cache lines and the \basesf character entries are distributed evenly, so $\prob[C\!=\!c]=\frac{1}{2}$ and $\prob[B\!=\!b\,|\,C\!=\!c]=\frac{1}{32}$, we see the maximum possible leakage value of $I(B,C)=1$, which means that we learn one bit of each \basesf character by observing the accessed cache line. If the table is not evenly distributed, the entropy for the sparser cache line decreases, making it easier to infer the respective \basesf character; however, at the same time, the entropy for the denser cache line increases, making up for an overall smaller leakage. If the alignment is not at a cache line boundary, but within a cache line, the table may spread over more than two cache lines, leading to a potentially higher leakage. In our experiments, we mostly observed 64 byte and 32 byte alignments, except for libraries compiled with the SGX framework: Due to the memory constraints, the standard Makefiles enable optimization for space (\lstinline{-Os} in GCC), which reduces the table alignment down to 1 byte. 
While the leaked information per \basesf character is rather small and capped at one bit, the redundancy imposed by storing multiple secret key parameters makes up for this, as we show in Section \ref{sec:key_recovery}.

\noindent
{\renewcommand{\arraystretch}{1.2}
\begin{table*}[btph]
	\caption{LUT properties and expected leakage of \basesf decoding implementations of several standard and SGX crypto libraries. The observed LUT alignment is taken from our test system and may vary between systems and package sources. The estimated leakage $\bm{I(B,C)}$ depends on the LUT size, its observed alignment and the distribution of relevant entries over cache lines.}
	\label{tab:crypt_libraries_overview}

    \centering
\begin{threeparttable}
	\begin{tabular}{l c c c c c c c c }
		Library & Version & Decode iterations & LUT size & LUT alignment (observed) & \# Cache Lines & $I(B,C)$\\
		\toprule
		
		\botan \cite{botangit} & 2.17.0 & 1 & 256 byte & variable (32 byte) & 5 & 0.974 bit\\ 
		
		\gnunettle \cite{gnunettlegit} & 3.6 & 1 & 256 byte & variable (32 byte) & 5 & 0.974 bit\\ 

		\mbedtls \cite{mbedtlsgit} & 2.24.0 & 2 & 128 byte & variable (32 byte) & 3 & 0.974 bit\\ 
		
		MS CryptoAPI \cite{cryptoapi} & 10.0.18362.476 & 1 & 80 byte & unknown\tnote{1}\ \ (64 byte) & 2 & 0.811 bit\\
		
		\nss \cite{nssgit} & 3.58 & 1 & 256 byte & variable (64 byte) & 4 & 0.696 bit\\
		
		\openssl \cite{opensslgit} & 1.1.1h & 2 & 128 byte & variable (32 byte) & 3 & 0.974 bit\\ 
		
		\rustsgx \cite{rustsgxgit} & 1.1.3\tnote{2} & 1 & 256 byte & variable (20 byte) & 5 & 0.564 bit\\ 
		
		\wolfssl \cite{wolfsslgit} & 4.5.0 & 1 & 80 byte & variable (64 byte) & 2 & 0.811 bit \\ 

		\bottomrule 
	\end{tabular}
	\begin{tablenotes}
	    \item[1] The source code is not publicly available, so we could not determine whether Microsoft uses a fixed or a variable alignment.
	    \item[2] The \basesf decoder itself is included in a separate package (version 0.13.0), which gets pulled into the SGX enclave.
	\end{tablenotes}
\end{threeparttable}
\end{table*}
}

\textit{Non-LUT-based \basesf decoding: }
Another approach for \basesf decoding is treating each case separately: Most characters (letters and numbers) are ASCII-encoded in contiguous chunks, with only few exceptions. Thus, one can test whether the current character is in a specific interval, and then simply add/subtract a certain constant which then yields the associated 6-bit value. This approach has, e.\,g., been used by \boringssl~\cite{googleboringssl} and the Rust \basesf package, although the latter has since moved to a \ac{LUT}-based implementation.

Depending on the binary layout of the code handling each case, an attacker may be able to acquire much more fine-grained information about each character than in a \ac{LUT}-based attack: If they can distinguish each case, which may be possible by counting the number of executed instructions per loop iteration, they learn whether the current character is an upper- or lower-case letter, a number, or a special symbol. This corresponds to more than 1 bit of information, even higher than the leakage induced by \ac{LUT}-based decoding.

\subsubsection{Exploiting the DER Format}
\label{subsubsec:exploiting_the_der_format}
Even though the majority of the detected leakages are found in the \basesf decoder, we also identified subtle secret-dependent computations in the DER decoder and the big number initialization. In DER, the parameters are not stored directly next to each other, but have a prefix denoting their type (integer, \lstinline{02}) and byte length (see Listing~\ref{lst:rsa_key_der} in the appendix). Since \basesf encoding divides the payload into 6-bit chunks, some chunks may contain bits from both a secret parameter and a byte belonging to DER formatting. If this DER byte is known to an attacker, they can reduce the remaining uncertainty from detecting the corresponding \ac{LUT} cache line, and infer up to 4 bits of the first or last secret parameter byte. While the parameter type byte is constant, the length byte is not; however, an attacker can learn the length of the parameter through other leakages, like in cases where a parameter is copied when initializing a big number object: In order to speed up arithmetic operations, many big number implementations divide their state into 64-bit integer chunks, which are initialized by copying the number bytes using bitwise operations like shifts and OR. The attacker can then simply count the number of loop iterations and thus learn the parameter length, if the loop is not constant-time.

\section{Cache attack on Intel SGX Enclave}
\label{sec:cache_attack_in_intel_sgx}
Attacking a simple lookup procedure, which mainly involves memory loads executed in a very short time frame, requires a high temporal attack resolution or a slowed down victim process. 
Thus, we attack the \basesf decoding process of RSA keys in an Intel \ac{SGX} enclave, which 
allows us to analyze the decoding process on a per-instruction basis. The attack we implemented is specific to the way \openssl implements the decoding of \basesf keys into its internal data format, especially the offline analysis part which leverages \openssl's access pattern to the \ac{LUT}. However, the translation from \basesf to binary by means of a \ac{LUT} is a recurring pattern in all of the libraries shown in Table \ref{tab:crypt_libraries_overview}. Thus, the general attack scheme is applicable to other libraries as well.

In short, our attack on the \basesf \ac{LUT}-based decoding process of RSA keys consists of several steps. First, we run OpenSSL's 
key decoding in an SGX enclave 
and execute it in a controlled, single-stepped fashion, where the corresponding memory page accesses to the \ac{LUT} and decoding function are tracked. We combine the page access monitoring with a classic \PrimeProbe attack on the \ac{LLC} to track which cache line of the decoding table is accessed when the investigated code is executed. The resulting trace is then processed during an offline analysis step, which outputs a cache line access pattern with the same length as the original \basesf string in the \ac{PEM} file holding the private key.

We first run the attack without mitigation against recent transient execution attacks like \ac{LVI} \cite{van2020lvi} and obtain mostly negative results. However, as we show in this section, running the same experiments with enabled mitigation drastically reduces noise in the measurements, which allows to reliably extract all information introduced by non-constant time behavior in the \basesf decoding.

\subsection{Attack description}

\subsubsection{Attacker model}
Intel \ac{SGX} aims to protect programs by running them in enclaves isolated by special hardware mechanisms. Ultimately, it allows enclaves to be guarded from a malicious \ac{OS} and otherwise rogue software environments and system administrators as long as the authenticity and integrity of the enclave and \ac{SGX} instance are verified by attestation~\cite{anati2013innovative, mckeen2013innovative, costan2016intel}. Consequently, attacking a process running in a protected enclave assumes an attacker with system level privileges having full control over the \ac{OS} kernel and the system BIOS: They are capable of translating virtual to physical addresses, manipulating page access bits and setting timed interrupts using the APIC timer. Additionally, they have access to the program's binary and control the unprotected application part. By using the \textit{SGX-Step} framework \cite{vanbulck2017SGXStep}, the enclave can be single-stepped.

\subsubsection{Cache Attack}

For our cache attack, we use \PrimeProbe with eviction sets.
After the discovery of \textit{Foreshadow}~\cite{van2018foreshadow}, Intel published a microcode fix which conducts an L1 cache flush on every enclave exit, so we are restricted to attacking the L3 cache.

\subsubsection{Attack process}
\label{subsubsection:attack_process}
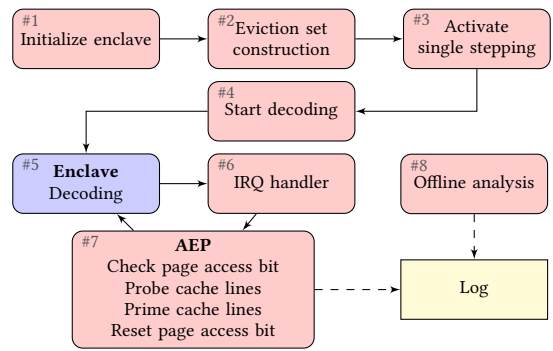
\begin{figure}[h]
    \centering
	\tikzstyle{block_victim} = [rectangle, draw, fill=blue!20, 
    text width=5.5em, text centered, rounded corners, minimum height=2.5em]
\tikzstyle{block_attacker} = [rectangle, draw, fill=red!20, 
    text width=5.5em, text centered, rounded corners, minimum height=2.5em]
\tikzstyle{rec} = [rectangle, draw, fill=yellow!20, 
    text width=5.5em, text centered, minimum height=2.5em]
\tikzstyle{line} = [draw, -latex']
\tikzstyle{every node}=[font=\footnotesize]

\begin{tikzpicture}[auto]
	\node [block_attacker, label={[shift={(3.5ex,-2.5ex)}, customdarkgray]north west:\#1}] (init_enclave) {Initialize enclave};
	\node [block_attacker, right=2em of init_enclave, label={[shift={(3.5ex,-2.5ex)}, customdarkgray]north west:\#2}] (construct_eviction_sets) {Eviction set construction};
	\node [block_attacker, right=2em of construct_eviction_sets, label={[shift={(3.5ex,-2.5ex)}, customdarkgray]north west:\#3}] (activate_single_stepping) {Activate\\single stepping};
	\node [block_attacker, below=0.5em of construct_eviction_sets, , label={[shift={(3.5ex,-2.5ex)}, customdarkgray]north west:\#4}] (start_decoding) {Start decoding};
	\node [block_attacker, below=0.5em of start_decoding, label={[shift={(3.5ex,-2.5ex)}, customdarkgray]north west:\#6}] (irq_handler) {IRQ handler};
	\node [block_attacker, below left=0.7em and -4.5em of irq_handler, label={[shift={(3.5ex,-2.5ex)}, customdarkgray]north west:\#7}, text width=9.5em] (aep) {\textbf{AEP}\\ Check page access bit\\ Probe cache lines\\ Prime cache lines\\ Reset page access bit};
	\node [block_victim, left=2em of irq_handler, label={[shift={(3.5ex,-2.5ex)}, customdarkgray]north west:\#5}] (decoding) {\textbf{Enclave}\\Decoding};
	\node [rec, right=3.6em of aep] (file) {Log};
	\node [block_attacker, above=1.9em of file, , label={[shift={(3.5ex,-2.5ex)}, customdarkgray]north west:\#8}] (offline_analysis) {Offline analysis};
	
	\path [line] (init_enclave) -- (construct_eviction_sets);
	\path [line] (construct_eviction_sets) -- (activate_single_stepping);
	\path [line] (activate_single_stepping) |- (start_decoding);
	\path [line] (start_decoding) -| (decoding);
	
	\path [line] (decoding) -- (irq_handler);
	\path [line] (irq_handler) -- (aep);
	\path [line] (aep) -- (decoding);

	\path [line, dashed] (aep) -- (file);
	\path [line, dashed] (offline_analysis) -- (file);
\end{tikzpicture}
	\caption{Attack process. Red: Attacker activity, Blue: Victim activity;
	The attacker is in control of the environment as well as the enclave host application and calls the victim code to start the attack process.}
	\label{tikz:attack_process_overview}
\end{figure}

Figure \ref{tikz:attack_process_overview} shows an overview of the attack process on \basesf decoding in Intel \ac{SGX}. 
We start with initializing the victim's enclave (\#1) and constructing eviction sets (\#2) for every cache set possibly containing the cache lines holding the \acl{LUT}. Since the \acl{LLC} is divided into slices, the number of required eviction sets is determined by the number of cache lines occupied by the \ac{LUT} times the number of slices. To construct the eviction sets, we implement an algorithm similar to the procedure presented by Liu et al. \cite{liu2015last} using virtual to physical address translation. 
After constructing the eviction sets, we use \textit{SGX-Step} to configure APIC timer interrupts which allow us to single step code running in the enclave  (\#3) and subsequently trigger the \basesf decoding (\#4). 

Next, we enter the enclave with the \texttt{EENTER} instruction and execute one instruction (\#5) during which the APIC timer interrupt arrives. The interrupt causes an \texttt{EEXIT}, which is followed by the IRQ handler (\#6) which redirects to our customized \ac{AEP} function (\#7). The latter is used to implement the attack code and finally resume the enclave, returning to state \#5. The cycle is terminated when the end of the \basesf procedure is detected.

Single stepping the victim code allows us to analyze the cache behavior on a per instruction resolution, and, since the enclave is in an interrupted state, our attack code in step \#7 is not time constrained. 
However, entering the enclave takes substantially longer than executing the next victim instruction, which adds potential for noise accumulation in the \ac{LLC} cache from other processes on the system. 

To reduce unwanted side-effects and noise in the \ac{LLC}, the \ac{AEP} routine starts with checking the page access bits of the memory pages holding the \ac{LUT} and the decoding routine, and then immediately continues with probing the cache. 

After probing the cache, the results of the cache eviction measurements and the page access states are stored to disk for offline analysis (\#8). Before resuming the enclave, the cache sets of interest are primed, the page access bits belonging to the memory pages holding the \ac{LUT} and decoding routine are reset and the APIC timer is reprogrammed.

\subsubsection{Offline Analysis}
\label{subsubsec:offline_analysis}

The data collected in step \#7 is processed in an offline analysis after the measurement finished. It contains cache eviction time measurements and page access information, as explained in \ref{subsubsection:attack_process}, for every single-stepped instruction. In the following, a \textit{measurement} refers to all the data collected for one single stepped instruction. The goal of the offline analysis is identifying those instructions that read data from the \ac{LUT} and determining the respectively accessed cache lines. Finally, a trace of cache line accesses corresponding to the \basesf characters in the private key's \ac{PEM} file is constructed.

We first determine the median eviction times and corresponding standard deviations for each eviction set over all measurements, excluding those with observed memory page access to the pages holding the \ac{LUT} and decoding function. Those measurements which show an access to both pages are, with a few exceptions at the beginning of the trace, the ones corresponding to an actual lookup operation. The median eviction times and standard deviation serve as a basis to determine evictions in the measurements which represent actual lookup operations. Prefiltering the measurements using the page access information significantly reduces the chance for false positives, meaning measurements will not falsely be identified as \ac{LUT} hit if there is noise in non-relevant cache probings. We identify an access to a cache line containing \ac{LUT} information by having an eviction time which differs by two standard deviations.

As stated in Section \ref{subsubsec:leakages_in_base64_dec}, \openssl looks up each symbol in a \ac{PEM} file at least twice. Additionally, it parses the \ac{PEM} file in blocks of 64 symbols. The symbols which are at the boundaries of a block are even parsed three times as they are checked for white space and end of line characters. Thus, we see a clear access pattern to the \acl{LUT}, which can be used to eliminate remaining irrelevant elements from the beginning of the trace and match the two passes for every 64 byte block against each other. The last step also allows for error correction or filling up gaps.

In order to extract each key parameter, the trace needs to be partitioned according to the \ac{DER} format, by identifying parameter lengths and removing meta data: As mentioned in Section \ref{subsubsec:exploiting_the_der_format}, \openssl leaks the parameter length information in the \lstinline{BN_bin2bn} method, which iterates over every byte in the \ac{DER} binary, and converts the data to an internal array representation. It can be attacked in a similar manner as the lookup operation, except that the single-stepped \PrimeProbe attack must be run against the cache line holding the instruction which loads the next key byte. Counting the number of evictions and translating them to the iteration count determines the length of each parameter easily. 

\subsection{Experimental Evaluation}

In the following, we describe the experimental setup to conduct the single-stepped cache attack against \basesf decoding, and discuss our observed results. We show that a mitigation against an attack in the transient domain greatly simplifies the process of leaking information from the decoding operation.

\subsubsection{Setup}
For the evaluation of the attack and leakage extraction we evaluated two different enclaves and took measurements on three different CPUs. First, we crafted an enclave containing the relevant code parts for \basesf decoding from \openssl and ran experiments on an Intel i5-8259U processor with an 6144 kB inclusive L3 cache and 4 GB main memory. The cache has 12 ways and, as assessed in our experiments, 8 slices with 1024 sets each. 

Second, we conducted the same measurements on an enclave  which decodes a \basesf encoded private key using the intel-sgx-ssl~\cite{intel-sgx-ssl} library in version 1.1.1k, compiled with default settings. The intel-sgx-ssl project compiles and installs the trusted \openssl libraries with and without mitigation by default. We linked our enclave against intel-sgx-ssl with \texttt{MITIGATION-CVE-2020-0551} set to \texttt{LOAD}, \texttt{CF} and no mitigation and compared the results. The measurements with intel-sgx-ssl were run on an Intel Xeon E-2286M with 16384 kB inclusive L3 cache and 16 GB main memory and on an Intel i5-6400 with an 6144 kB inclusive L3 cache and 4 GB main memory.

All CPUs used the latest stable microcode patches. For compatibility with SGX-Step, we used Intel SGX SDK version 2.11. We disabled hardware prefetching for the L1 and L2 caches on each core. Additionally, the CPU frequency was fixed to the processor's base frequency on all cores, Intel Speedstep was disabled and the maximal C-State was set to 0 in order to decrease variability in the measurements.  Finally, we assigned the enclave and its host application to a specific logical core, which was removed from the \ac{OS} scheduler.

\subsubsection{Results}

The first experiments were run without configuring the \texttt{make} process to apply mitigations against \ac{LVI} \cite{van2020lvi}, which are available since the Intel \ac{SGX} \ac{PSW} and \ac{SDK} version \texttt{2.9.100.2}.

The results are very noisy and hardly exploitable, in fact most measurement runs are not usable at all, as the eviction time measurements of the \PrimeProbe attack are inconclusive: Extracting the sequence of lookups of \basesf symbols is not possible, as both monitored sets were accessed, even though we performed a single-stepping attack.
Simultaneous accesses are likely caused by speculative or out-of-order accesses of the lookups, as subsequent lookups are only few instructions apart. While single-stepped execution ensures that only one instruction \emph{commits} between interrupts, several are issued in parallel in that time window. We further suspect that this transient effect is amplified by resetting the page accessed bits, which increases the out-of-order window. The measurements for this experiment without the \ac{LVI} mitigation reveal that most of the time, evictions are observed for \emph{both} investigated cache sets, which renders a distinction infeasible. Table \ref{tab:cache_hits_in_both_sets_no_lvi_mitigation} shows a few example measurements for both cache sets without mitigation.

{\renewcommand{\arraystretch}{1.2}
\begin{table*}[t]
	\caption{Exemplary eviction set measurements during cache set probing from the experiments without \ac{LVI} mitigations. The rows show measurements with and without accesses to the \ac{LUT}, which correspond to accesses to the \ac{LUT} page. Each time we observe a page access, we also see an eviction (orange); however, those evictions always occur in both cache sets, not allowing us to draw any conclusions which set has been accessed. The last row specifies the slice number.}
	\label{tab:cache_hits_in_both_sets_no_lvi_mitigation}
	
\centering
	\begin{tabular}{l | r r r r r r r r | r r r r r r r r }
	Page Access & & & & Set 1 & & & & & & & & Set 2 & & & & \\
	\toprule
No & 848 & 784 & 756 &  808 & 842 & 780 & 758 & 852 & 888 & 800 & 794 & 760 & 798 & 852 & 760 & 788 \\
No & 846 & 778 & 750 &  806 & 844 & 778 & 756 & 842 & 888 & 802 & 782 & 760 & 804 & 848 & 760 & 788 \\
\color{black}{Yes} & \color{black}{842} & \color{black}{784} & \color{black}{754} & \color{orange}{2486} & \color{black}{846} & \color{black}{782} & \color{black}{756} & \color{black}{846} & \color{orange}{2478} & \color{black}{802} & \color{black}{790} & \color{black}{764} & \color{black}{802} & \color{black}{844} & \color{black}{764} & \color{black}{788} \\

\color{black}{Yes} & \color{black}{842} & \color{black}{778} & \color{black}{756} & \color{orange}{2488} & \color{black}{844} & \color{black}{784} & \color{black}{766} & \color{black}{850} & \color{orange}{2940} & \color{black}{802} & \color{black}{790} & \color{black}{766} & \color{black}{808} & \color{black}{846} & \color{black}{760} & \color{black}{784} \\

	\bottomrule
	 & S1 & S2 & S3 & S4 & S5 & S6 & S7 & S8 & S1 & S2 & S3 & S4 & S5 & S6 & S7 & S8
	\end{tabular}

\end{table*}
}

\begin{figure*}[h]
	\includegraphics[width=1\textwidth]{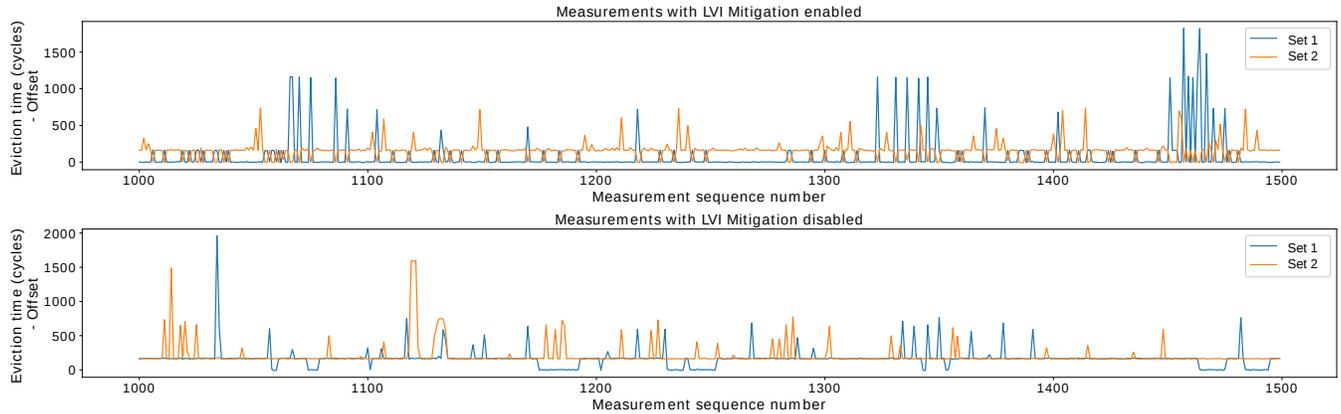}
	\caption{Eviction set measurements with and without \ac{LVI} mitigations enabled. The blue and orange lines correspond to the eviction times of the cache sets holding the \ac{LUT}. To level both graphs, the respective mean measurement time has been subtracted, resulting in an expected value of 0 when the corresponding set has not been accessed. In the upper plot, with enabled \ac{LVI} mitigations, we see a clear separation of both sets: If the orange graph is positive, the blue one is 0, and vice versa. Note that the \ac{LUT} entries are not evenly distributed, leading to a bias towards the orange set. In the lower plot, without \ac{LVI} mitigations, we see that most of the time both sets are hit, so a clear separation is impossible.}
	\label{fig:compare_time_series}
\end{figure*}

Next, we repeated the measurements with \texttt{MITIGATION-CVE-2020-0551} set to \texttt{LOAD}~\cite{intellviadvisory}. This \ac{LVI} mitigation places load fences after every instruction which has a load micro-op \cite{van2020lvi, intelsgxdevref}. Consequently, it prohibits out-of-order execution of instructions after the traced load instruction, which otherwise might have accessed further cache lines in the \ac{LLC}.
Figure~\ref{fig:compare_time_series}  depicts a comparison of eviction times for measurements with and without \ac{LVI} mitigation. It is observable that with the \ac{LVI} countermeasure, only one of the two monitored cache lines is accessed, while both are accessed when the countermeasure is turned off.
We thus conclude that the \ac{LVI} countermeasure greatly enhances granularity of cache attacks.

The attack we ran against \basesf decoding in Intel \ac{SGX} requires \emph{only one execution} to create a trace, which leaks all information we can obtain from priming and probing the cache sets holding the \ac{LUT}. 
In order to determine the reliability of the measurements, we ran the attack 100 times against the same key and tried to extract the respective cache line access trace. For our experiments, we aligned the \acl{LUT} on a 64 byte boundary, such that the \ac{LUT} used in \openssl spread over exactly 2 cache lines. The cache access trace created by the offline analysis is a string with elements from \{\texttt{1}, \texttt{2}, \texttt{x}\}, where \texttt{x} means that no clear distinction can be made and \texttt{1} and \texttt{2} identify the accessed cache lines. 

Finally, each of the extracted traces is checked for the correct length and compared against the actual key, by checking for each \basesf symbol whether it matches the cache line  access. The \ac{PEM} file holding the 1024 bit test key has a length of 848 \basesf symbols, thus requiring the same length for the measured cache access trace.

\begin{figure*}
	\centering
    \includegraphics[width=\textwidth]{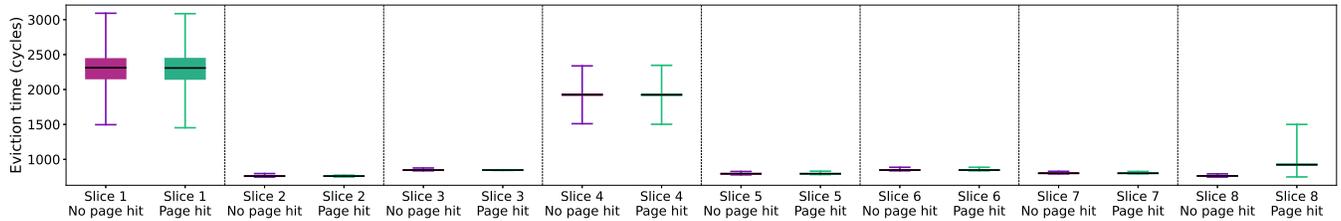}
	\caption{Eviction times measured after every instruction during the decoding process for all cache sets over all slices possibly holding cache line 2 of the \ac{LUT}. Violet boxes and whiskers show the eviction time of all measurements in which the memory pages of the \acl{LUT} and decoding function were not accessed. Green shows the measurements for which the observed pages were accessed. Since no knowledge of the slice mapping is assumed, the slice numbers cannot be matched to a logical CPU core and will be assigned differently in every execution. In the depicted case, the victim's accesses map to slice 8.
	}
	\label{fig:slice_eviction_times}
\end{figure*}

Figure \ref{fig:slice_eviction_times} depicts the eviction time measurements for all sets over all slices possibly holding cache line 2 of the \ac{LUT} when probing the corresponding eviction sets.

The sets in all slices but slice 8 reveal the same spectrum of eviction times for measurements with and without observed page accesses. However, for slice 8, a clear deviation in eviction time measurements can be observed, which allows the detection of \ac{LUT} accesses.

\begin{figure}[h]
	\includegraphics[width=1\columnwidth,height=17em,trim=2.5cm 0cm 3cm 0cm, clip]{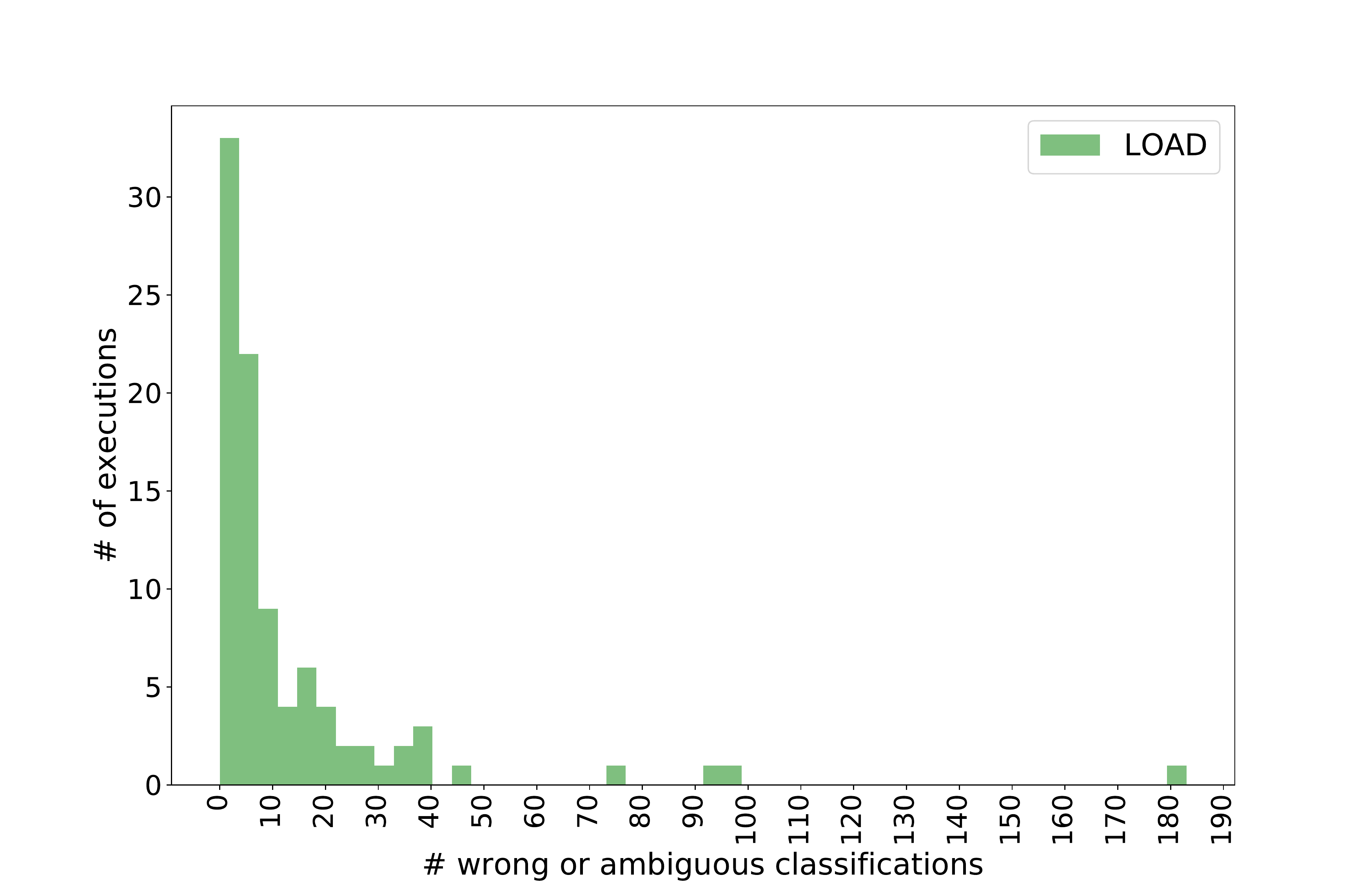}
		\caption{Number of ambiguous or wrong cache access measurements for 93 executions of the experiment, where we were able to recover the full trace. The histogram shows that the majority of executions had less than 10 measurements where the classification was wrong or ambiguous. This corresponds to around 1.2\% of the full trace. The measurements were performed on an Intel Xeon E-2286M against our OpenSSL-based enclave.}
	\label{fig:ambiguous_classifications}
\end{figure}

\paragraph{Self-Crafted Enclave} The histogram in Figure~\ref{fig:ambiguous_classifications} shows the number of trace elements which could not be classified (\texttt{x}) or which received a wrong classification per execution. The measurement was taken on the Intel i5-8259U with the ``self-crafted'' enclave and LVI mitigation level set to \texttt{LOAD}.

The data shows that the attack runs stable in most cases. In 93 of the measurements the automated offline analysis is able to extract a sequence of correct length, and in 62 the number of cache line accesses which could not be classified is less than 10, which is only 1.2\% of the full trace. Figure \ref{fig:ambiguous_classifications} shows that there are only few measurements with more than 30 ambiguous or wrong cache line classifications. Additionally, in none of the 93 measurements, for which extracting the sequence was feasible, a cache line hit was detected for the wrong cache line. 
This very reliable classification can partially be attributed to {\openssl} looking up each symbol twice.

For the key reconstruction, we are only interested in the lower half (least significant part) of bits of every parameter, as explained in Section \ref{sec:key_recovery}. This reduces the number of relevant missing cache line classifications to about the half. Moreover, due to a random distribution of missing information, running the attack twice is sufficient to obtain a (near) complete trace.

\begin{figure}[h]
	\includegraphics[width=1\columnwidth,height=17em,trim=2.5cm 0cm 3cm 0cm, clip]{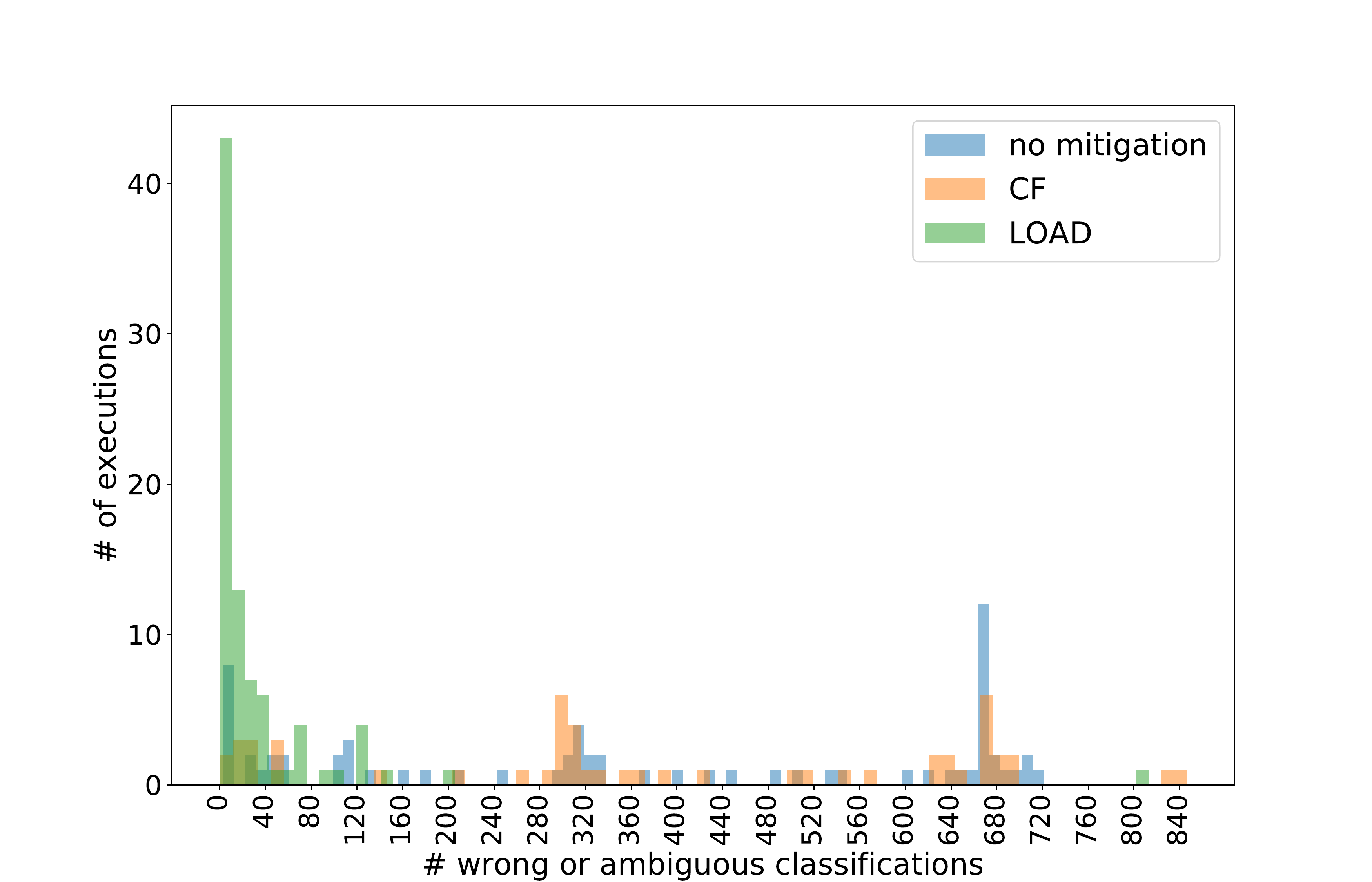}
		\caption{Number of ambiguous or wrong cache access measurements across different mitigation levels.  Traces with \texttt{LOAD} mitigation level contain considerably less errors. The measurements were performed on an Intel Xeon E-2286M against the intel-sgx-ssl enclave.}
	\label{fig:ambiguous_classifications_intl-sgx-ssl}
\end{figure}

\paragraph{Enclave with intel-sgx-ssl} In Figure~\ref{fig:ambiguous_classifications_intl-sgx-ssl}, the measurement results on the Intel Xeon E-2286M with an enclave using intel-sgx-ssl to decode the \basesf encoded key are depicted. The measurement was taken with \texttt{MITIGATION-CVE-2020-0551} set to \texttt{LOAD}, \texttt{CF} and without mitigation. The results with the mitigation level set to \texttt{LOAD} show that 14 of 100 traces don't have any errors and 42\% of the automatically extracted traces have less than 1.2\% of errors (10/848). The attack also works when no mitigations are applied, but significantly worse: On the Xeon, only about 7\% of the observed traces have less than 10 missing classifications and there is none without wrong or ambiguous trace elements. As leakage is already quite low, errors must be avoided at all cost, so many traces are required to obtain a reliable trace with no mitigations. The \texttt{CF} mitigation is comparable to no mitigations, as it does not inject fences after load instructions in the decoding routine, but only for control flow related instructions.

On the  Intel i5-6400, the results with no mitigations applied are better, but still clearly worse than with \texttt{LOAD} mitigations enabled. 

\subsubsection{Practical relevance of LVI mitigations}
Setting \texttt{MITIGATION-CVE-2020-0551} to \texttt{LOAD} has a high performance impact. In general, it is hard to say whether this mitigation is applied in commercial enclaves; that also holds for open source software, since the mitigations are activated by explictly setting an environment variable. However, we believe that this mitigation has its value in practical applications and that should be applied to secret-dependent workloads like key loading procedures and cryptographic operations. 

In general, Intel recommends applying the \texttt{MITIGATION-CVE-2020-0551} on LVI-affected platforms~\cite{intellvitechdoc}: "Intel SGX Attestation Service will report a new status code, \texttt{SW\_HARDENING\_NEEDED}, to indicate the platform is affected by a security advisory for which software hardening is recommended". Intel recommends enclave developers to "determine the level of software hardening that their environment requires, based on risk analysis and an evaluation of the performance impacts of mitigation".

The \texttt{CF} (Control-Flow-Mitigation) mitigation level will only protect against LVI gadgets which use control-flow instructions for secret transmission. However, secret transmissions with LVI can also be encoded into the data flow~\cite{van2020lvi}, so memory load instructions have to be protected with \texttt{LFENCE}s as well. Since this is rather important in secret-dependent algorithms, there is a practical relevance for the LOAD mitigation level in this case.

We found several concrete applications using these mitigations by default or offering a version with mitigations applied:
\begin{itemize}
    \item Inclavare Containers~\cite{inclavare-containers} and the RUST SGX SDK~\cite{rustsgxgit} enable their users to apply the mitigations. For the enclave-tls module of the former, it is even stated in the documentation that the SGX LVI mitigation is enabled by default, but not which level~\cite{inclavare-containers-lvi-mitigations}. Both frameworks consider both levels.
    \item SecretNetwork \cite{secretnetwork} enables the LOAD mitigation level in their deployment / Docker files \cite{secrectnetworkdeployment}.
    \item According to a GitHub issue~\cite{asylousagelvimiti}, Asylo~\cite{asylo} uses~\texttt{MITIGATION-CVE-2020-0551} set to \texttt{LOAD} by default since May 2020.
    \item Intel offers a variant of its Crypto API Toolkit with all mitigations enabled~\cite{crypto-api-toolkit-lvi-mitigations}. Additionally, as already stated, by default intel-sgx-ssl builds versions for each mitigation level.
\end{itemize}

\section{RSA Key Recovery}
\label{sec:key_recovery}

In the following, we will adapt the algorithm of Heninger and Shacham~\cite{heninger2009reconstructing} to the
setting, where only information about certain blocks of bits is known.
Here, we only give a high-level overview and refer the
reader to the appendix, which contains a complete formal description of both the setting and the algorithm.

We first formalize the setting, describe the adapted algorithm, and analyze its
running time.
Finally, we discuss optimizations used in our implementation.

\subsubsection*{Blockwise Knowledge}
We consider the situation that some blockwise knowledge about the secret key
$\sk^{\star}=(p^{\star},q^{\star},d^{\star},d^{\star}_{p},d^{\star}_{q},{q^{\star}_{p}}^{-1})$
was obtained.
In the following, we focus on the first five variables and treat $\sk^{\star}$
as a quintuple on the variables $\Vars=\{p,q,d,d_{p},d_{q}\}$.
To simplify the notation, for $v\in \Vars$, we denote the corresponding entry in
some key $\sk$ by $\sk[v]$.
We show in Sec.~\ref{app:last} that integrating the last variable $q_{p}^{-1}$
into the key-recovery approach does not directly give a usable linear equation
in contrast to the other variables.

From a high-level perspective, our attack gives us the following information: 
For each \emph{6-bit block} of a variable $v\in \Vars$, we know that this
block belongs to a certain \emph{cache line}. 
This knowledge allows us to rule out the values of the other cache lines for
this block.
For example, we might know that the first $6$ bits of $p^{\star}$ belong to cache line~$i$. 
As we also know the content $C_{i}\subseteq \{0,\ldots,2^{6}-1\}$ of cache
line~$i$, we can reduce our search space for these $6$ bits from the complete
space $\{0,\ldots,2^{6}-1\}$ down to $C_{i}$, but we still
have a remaining uncertainty about which concrete value in $C_{i}$ was used.
In contrast, in the scenarios studied
in~\cite{DBLP:conf/asiacrypt/PatersonPS12,DBLP:conf/crypto/HeneckaMM10,heninger2009reconstructing},
the knowledge always was about \emph{single} bits.
Hence, the attacks here might have given the information that the fifth bit of
$p$ equals $0$.
The uncertainty in this scenario comes from the fact that this information could
potentially be wrong (e.\,g.~due to a bit-flip in the cold-boot scenario).

\subsubsection*{Modeling the Scenario}
As described above, in the situation given by our attack, we do not have
observations on single bits, but on blocks consisting of $6$ bits, the length of
a \basesf symbol.
To generalize this knowledge, we let $b\in \mathbb{Z}_{> 0}$ be the
\emph{blocksize}, i.\,e.~the length of the block on which we have obtained our
knowledge.
For a variable~$v$, we denote the $j$-th block of length $b$ as $\block_{j}(v)$,
e.\,g.~the six least significant bits of $p$ are denoted as $\block_{0}(p)$. 
In our attack, we make use of the fact that the possible values for
$\block_{j}(v)$ are partitioned into different sets to model the different cache
lines used in our attack.
To formalize this, we consider a partition $\parts=(\parts_{1},\ldots,\parts_{|\parts|})$ of all possible $b$-bit values
$\{0,\ldots,2^{b}-1\}$.
The set $\parts_{i}$ would thus correspond to the content of $C_{i}$ of cache
line $i$.
Our algorithm is now given an \emph{observation} about a certain key $\sk^{\star}$ stating that for each variable
$v\in \Vars$, the block $\block_{j}(v)$ belongs to $\parts_{i}$.
In our concrete application, this translates to the knowledge that the $j$-th 
\basesf symbol  of variable $v$ belongs to  cache line $i$. 

\subsection{Recovery Algorithm}
The main idea of the algorithm is to reconstruct the different bits of the
secret key $\sk^{\star}$ iteratively.
We build up a set of \emph{candidates} iteratively. 
Each such candidate is a guess for the least significant bits of the true secret key $\sk^{\star}$ compatible with our observation and the RSA equations. 
We start our algorithm by producing a single candidate $\mywidetilde{\sk}$ of depth~$1$, i.\,e.~each variable only consists of a single bit.
Informally, the depth of a candidate is the number of bits each variable has. 
We then apply the $\mathsf{expand}$ operation on $\mywidetilde{\sk}$ to obtain two candidates $\mywidetilde{\sk}_1$ and $\mywidetilde{\sk}_2$ of depth $2$ by using the RSA equations described by Heninger and Shacham~\cite{heninger2009reconstructing}.
Whenever a candidate has reached depth of a multiple of $b$, i.\,e., $j\cdot b$ for some $j$, we apply the $\mathsf{check}$ operation on this candidate to verify that the last produced block $\block_j(v)$ of each variable $v$ is possible under our observation. 
If this candidate does not fit to our observation, we prune it.
We repeat these operations until a target depth~$D$ is reached.
All produced candidates of depth~$D$ are output. 
This target depth is chosen such that the remaining bits can be reconstructed via the Coppersmith method~\cite{DBLP:journals/joc/Coppersmith97, DBLP:conf/ima/Howgrave-Graham97,DBLP:phd/de/May2003}. 

Our algorithm first performs these operations in a breadth-first fashion to utilize parallelisation and then in a depth-first fashion (see 
Figure~\ref{fig:algorithm}).
The $\mathsf{expand}$ operation uses a set of $4$ modular equations on $5$ variables and the $\mathsf{check}$ operation compares the generated candidates to our observations.

\subsection*{Pseudocode of the key-reconstruction algorithm}
\begin{figure}[h]
  \centering
  \begin{mdframed}[backgroundcolor=verylightgray]
  \pseudocode[mode=text,linenumbering,head={Input: Observation $\obs(\parts)$, target
    depth $D$}]{
    find valid triples $(k,k_{p},k_{q})$\\
    \pcfor each possible triple $(k,k_{p},k_{q})$:\\
    \t initialize empty stack $S$\\
    \t add initial candidate $\mywidetilde{\sk}(k,k_{p},k_{q})$ to $S$\\
    \t \pcwhile $S$ is not empty:\\
    \t[2] let $\mywidetilde{\sk}=S.\text{pop}()$\\
    \t[2] let $\mywidetilde{\sk}_{1},\mywidetilde{\sk}_2=\mathsf{expand}(\mywidetilde{\sk})$\\
    \t[3] \pcfor $\beta\in \{1,2\}$:\\
    \t[4] \pcif $\text{depth}(\mywidetilde{\sk}_{\beta}) \geq D$: \textbf{output} $\mywidetilde{\sk}_{\beta}$\\
    \t[4] \pcif $\text{depth}(\mywidetilde{\sk}_{\beta}) \bmod b = 0$ and
      $\mathsf{check}(\obs,\mywidetilde{\sk}_{\beta})$:\\
    \t[5] $S.\text{push}(\mywidetilde{\sk}_{\beta})$\\
    \t[4] \pcelse: $S.\text{push}(\mywidetilde{\sk}_{\beta})$
  }
\end{mdframed}
  
  \caption{Concise description of our adapted key-reconstruction algorithm}
  \label{fig:algorithm}
\end{figure}

\subsection{Analyzing the Algorithm}
In the following, we analyze the number of candidates of depth~$i$, produced by
the algorithm.
To do so, we need some probability notions. 
Let $\pr=(\pr[1],\ldots,\pr[k])$ be a \emph{probability vector} of length~$k$, i.\,e.~$\pr\in
[0,1]^{k}$ with $\sum_{i=1}^{k}\pr[i]=1$.
For $\alpha \geq 0$ with $\alpha\neq 1$, the  \emph{R\'{e}nyi entropy}
$H_{\alpha}(\pr)$ measures the amount of information given by $\pr$ and is
defined as
$ H_{\alpha}(\pr)=\frac{\alpha}{1-\alpha}\log\left( \sum_{i=1}^{k}\pr[i]^{\alpha} \right)$.

The special case for $\alpha=2$ is called the \emph{collision entropy}, which we will need in the run time analysis of our algorithm, similar to~\cite{DBLP:conf/ches/BernsteinBGBHLV17}. 
Intuitively, the usual Shannon-entropy used in Table~\ref{tab:crypt_libraries_overview} gives the complete amount of information available, but we can only use certain events to discard candidates not belonging to the observed cache line, namely non-collision events. 

To simplify the analysis of our algorithm, we use the heuristical assumptions
of~\cite{heninger2009reconstructing,DBLP:conf/ches/BernsteinBGBHLV17}, namely
\begin{assumption}\ 
  \begin{enumerate}
  \item Upon random choice of $\sk^{\star}$, for each $v\in \{p,q,d,d_{p},d_{q}\}$ and
    each block $\block_{j}(v)$, we have
    $\Pr[\block_{j}(v)\in \parts_{j}]=\frac{|\parts_{j}|}{2^{b}}$ and these
    probabilities are independent.
  \item Once a bit in a candidate $\mywidetilde{\sk}$ is set incorrectly (w.\,r.\,t.~$\sk^{\star}$), the set of
    satisfying solutions to the four congruences behaves randomly and
    independently. 
  \end{enumerate}
\end{assumption}

Using these, we can bound the expected number of candidates.

\begin{theorem}
  \label{thm:growth}
  Let $C$ be a set of incorrect candidates with depth $j\cdot b$. 
  After expanding these candidates $b$ times, the expected number of incorrect
  candidates after pruning 
  is  $|C|\cdot 2^{b- 5\cdot H_2(\pr)}+2^{b}-1$, where
  $\pr[i]=|\parts_{i}|/2^{b}$. 
\end{theorem}

It is easy to see that we have exactly $2^{b-1}$ initial candidates of depth $b$.
We can thus conclude the following theorem about the expected number of
candidates.

\begin{theorem}
  \label{thm:final}
  The expected number of incorrect candidates with depth $j\cdot b$ is at most 
$ 2^b\cdot \sum_{i=0}^{j}(2^{b- 5\cdot H_2(\pr)})^{i} = 2^b\cdot \frac{(2^{b- 5\cdot H_2(\pr)})^{j+1}-1}{2^{b- 5\cdot H_2(\pr)}-1}$.
\end{theorem}

\subsection{Termination of the Algorithm}
Finally, we need to describe how to set the target depth $D$ of our algorithm.
In a naive approach, we could set $D=\enc{\sk^{\star}[p]}$ and then test for all candidates $\mywidetilde{\sk}$
of depth $D$, whether $\mywidetilde{\sk}[p]$ is a factor of $N$.
But using a lattice-based approach, we can factor $N$ much faster.
The algorithm of Boneh, Durfee, and Frankel shows that it is sufficient to obtain
$\enc{\sk^{\star}[p]}/2$ bits of $p$ to factor~$N$ in polynomial time~\cite[Corollary~1]{DBLP:conf/asiacrypt/BonehDF98}.
By setting our target depth $D=\enc{N}/4$ and using the algorithm of Boneh, Durfee, and Frankel on
all candidates $\mywidetilde{\sk}[p]$ output by our algorithm, we can reconstruct the
correct secret key $\sk^{\star}$.
Together with Theorem~\ref{thm:final}, this shows that $(2^{b- 5\cdot H_2(\pr)})^{\enc{N}/(4b)}$ calls to the lattice algorithm are
sufficient to reconstruct $\sk^{\star}$.
Table~\ref{tab:entropy} contains the total number of calls to the lattice algorithms
for different collision entropies for blocksize $b=6$ (as in our attack),  compared with the security level in bits. 

\begin{table}
  \caption{Overview on the number of calls to the lattice algorithm for blocksize $\bm{b=6}$ compared with the security level of the key length. Here, B denotes bits and L the number of calls to the lattice algorithm.}
  \label{tab:entropy}
  \centering
  \begin{tabular}{ccccc}
    \toprule
    Key (B) & level (B) & $H_{2}=1$ (L) & $H_{2}=1{.}1$ (L) & $H_{2}=1{.}15$ (L) \\
    \midrule
    $1024$ & $2^{80}$ & $2^{49}$ & $2^{29}$ & $2^{19}$\\
    $2048$& $2^{112}$ & $2^{92}$ & $2^{50}$ & $2^{29}$\\
    \bottomrule
  \end{tabular}
\end{table}

\subsection{Experimental Evaluation}
To make the connection between the algorithm described above and our attack more explicit, the partition $\parts$ corresponds to the (usually two) different cache lines and the block length $b$ is $6$ due to the \basesf encoding. 
To reconstruct the RSA key completely, 
we implemented the adapted algorithm in C++ and implemented the final reconstruction step via the lattice algorithm \texttt{small\_roots} in \texttt{Sagemath 9{.}0}. 
Note that, due to the depth-first approach used in the algorithm, it is highly
parallelizable: if we are given $K+L$ processors, we can compute the
first $K$ candidates in a breadth-first fashion, distribute them across the
processors, and run them in depth-first fashion.
The remaining $L$ processors can then be used to apply the lattice algorithm on
all candidates of  length $D$. 

Table~\ref{tab:experiments} contains the experimental results of our algorithm for different key lengths.
In order to obtain these experimental results, we used idealized inputs to our algorithm, which were generated by hand, and represent a separate trace for every parameter.
\begin{table}
 \caption{Experimental Evaluation of our implementation on different key lengths and different cache distributions. }
  \label{tab:experiments}
  \centering
\begin{tabular}{ccrrr}
     \toprule
     Length &  Cache Dist. & \#Cand. & $\mathsf{genCands}$ [s] & $\mathsf{testCand}$ [s] \\
     \midrule
     256 & 38/26 &795{,}712 & 31 & 52{,}251 \\
     256 & 32/32 &31{,}760 & 2 & 1{,}740\\
     512 & 32/32& 2.08$\cdot10^8$ & 1.03$\cdot10^5$ & 1.53$\cdot10^7$ \\
     \bottomrule
\end{tabular}

\end{table}
Our experimental results showed that the running time of $\mathsf{genCands}$ and $\mathsf{testCand}$ is relatively stable per candidate with at most $0{.}0006$ seconds for $\mathsf{genCands}$ and at most  $0{.}07$ seconds for $\mathsf{testCand}$. 
An extrapolation shows that such a non-optimized implementation does not yet give an algorithm that reconstructs the complete $1024$-RSA key within a week: The generation of all candidates via $\mathsf{genCands}$ would take about $60$ CPU years and the reconstruction via $\mathsf{testCand}$ would take about $6{,}000$ CPU years.
We estimate a cost of about $1{,}000{,}000$ dollars on AWS and accordingly a few $100{,}000$ dollars on cheaper bare-bone clouds if we simply use many copies of \texttt{small\_roots}. 
But, as shown by the evolution around the Data Encryption Standard (DES), the time to brute-force over a search space of $2^{50}$ (as given in our case for $1024$-RSA) can be drastically reduced by more specialized hardware.
More concretely, \cite{DBLP:conf/ches/KumarPPPS06} uses 120 low-cost FPGAs and can make about $5\cdot 10^{10}$ DES calls per second allowing to break DES within two weeks. 
We thus expect more specialized hardware will lead to a  reconstruction time of a few weeks.

\section{Mitigations}

The demonstrated attack and library analysis show that not only cryptographic implementations themselves need to be protected against attacks, but that it is equally important to shield utility functions from side- and controlled-channel attacks, if they process secret data. We propose two mitigations for the \basesf attacks described in this work: First, we describe a constant-time variant of the original \acl{LUT}-based decoding algorithm, and discuss the constant-time case decision approach from {\boringssl}~\cite{googleboringssl}. Additionally, we highlight how adjusting existing best practices for key storage can help to reduce the surface for attacks on utility functions in general.

\subsection{Constant-time decoding}
\subsubsection{LUT-based}

A naive mitigation to our attack on the \acl{LUT} would work as follows: In order to make sure that the decoding of each symbol happens in constant time, each entry of the \acl{LUT} is accessed for each decoded symbol, and the correct symbol is selected using a mask. This approach will decrease decoding performance drastically, since decoding of each symbol does require 128 lookups (the size of the \ac{LUT} in bytes), instead of only one.

\begin{figure}[h]
\begin{mdframed}[backgroundcolor=verylightgray]
\begin{lstlisting}[basicstyle=\scriptsize\ttfamily,language=C++]
__attribute__ ((aligned (64)))
unsigned char lut[128] = { ... };

uint8_t decode_aligned(unsigned char b64ch) {
  uint8_t result = 0x00, mask = 0xAA;
  
  unsigned char idx[2] = { b64ch 
  
  for(unsigned char i = 0; i < 2; ++i) {
    mask = 0xFF ^ ((idx[i] == b64ch) - 1);
    result = result | (lut[idx[i]] & mask);
  }
  return result;
}
\end{lstlisting}
\end{mdframed}
\caption{Optimized constant-time decoding of a single \basesf character, with a 64-byte aligned lookup table. Note that the LUT only spans two cache lines, so two accesses are sufficient in our leakage model.} \label{lst:mitigation_aligned}
\end{figure}

To improve the performance of our naive mitigation, we add a constraint on the memory alignment as shown in Listing \ref{lst:mitigation_aligned}. By instructing the compiler to align the \ac{LUT} to 64 byte, it is only necessary to access each line once per symbol, which ensures that a controlled-channel attacker cannot determine the correct access in our leakage model. Therefore, we always access the \ac{LUT} at \texttt{b64ch mod 64} and \texttt{(b64ch mod 64) + 64} and select the correct lookup with a mask as before. In case the current index is smaller than 64, the first access correctly decodes the symbol, otherwise the second. The overhead of the \ac{LUT} dummy access should be negligible, compared to operations like asymmetric decryption.

\subsubsection{Case decision-based}
Google's \boringssl \cite{googleboringssl} already implements a constant-time \basesf decoding approach.

Constant-time behavior in \basesf decoding is achieved by a \ac{LUT}-free implementation. In a first step, it is determined to which part of the ASCII table the currently decoded symbol belongs. Then, the corresponding binary value is selected using a mask. Listing \ref{lst:constant_time_boringssl} in the appendix shows the relevant part of the decoding routine from \boringssl. Other examples for constant-time case decision-based \basesf decoding are libsodium~\cite{libsodiumgit} and Nimbus-JOSE-JWT~\cite{nimbus-jose-jwt}.

We believe that using a case decision-based approach has some advantages over using a LUT: Most cryptographic libraries already offer well-tested and portable macros for constant-time comparison and selection, so employing a separate technique in utility functions does not make much sense. Also, the LUT-based technique highlighted in Listing \ref{lst:mitigation_aligned} still makes certain assumptions on the underlying hardware and leakage behavior, which may not apply when compiling the same code for different target platforms. Finally, due to the relatively few calls the performance difference is negligible. We thus recommend to consider replacing LUT-based decoding functions by case decision-based implementations.

\subsection{Key Storage Practices}
Our \basesf decoding attack against RSA keys can also be mitigated by using encrypted PEM files: In this case, the attacker would only learn parts of the ciphertext, and is not able to derive the contained key. After \basesf decoding, the DER-encoded key is decrypted and decoded. Assuming that the key loading routine uses the same symmetric primitives that a given crypto library offers anyway (and which are thus subject to thorough security evaluation), this method reduces the attack surface to the key decoder and the key instantiation in memory. However, from our experience, many default server configurations use unencrypted key files, especially when doing automated replacement of keys and certificates (e.\,g., Let's Encrypt). Since these files are usually bound to a single system or instance, and do not leave this environment, a compromise of such a file would almost certainly also mean a compromise of the entire system and thus the passphrase needed to decrypt the key. On the other hand, if those private key files are not intended to be transmitted over the network or stored in text-based configuration files, there is no real benefit in using \basesf at all, since it just adds overhead and increases the attack surface: In such cases, simply storing the binary DER data would be sufficient.

\section{Related Work}

In general, leakage in key decoding is not a new concept: The authors of DATA~\cite{WeiserDATA} briefly mention true positives in {\openssl}'s key loading functionality, but did not further investigate the issue. In ~\cite{garcia2020certified}, the authors use alternative, but mathematically equivalent key representations to trigger specific non-hardened branches of the decoding routines, which deal with less common key formats and have thus been overlooked in prior research. However, they do not target generic utility functions, but arithmetic aspects of key decoding. To the best of our knowledge, the only other attack targeting utility functions is Medusa \cite{moghimi2020medusa}.
Medusa is an attack which leaks key information during \basesf decoding in \openssl. However, their focus is on extracting information from the transient domain and attacking the associated \texttt{rep mov} instruction. Such attacks are only possible if SMT is enabled and if the SGX enclave shares the core with a malicious process. Intel advises against such operations~\cite{MDS}. Our attack, however, does not need simultaneous access to the neighboring vCores and works fine on enclaves with disabled hyperthreading.
Furthermore, we do not only concentrate on a single instruction, but 
present a systematic analysis of key decoding functionality in several widely  used cryptographic libraries and show that these utility functions leak sensitive information despite and because of the mitigation introduced due to other microarchitectural attacks on \ac{SGX} and through the transient domain \cite{lipp2018meltdown, kocher2019spectre, SGXpectre18, van2018foreshadow, van2020lvi}. We leverage techniques common in the microarchitectural attack domain like \PrimeProbe and combine them with a recent attack framework~\cite{vanbulck2017SGXStep} to extract all available leakage introduced through non-constant time behaviour of the \basesf decoding process and analyze the leakage with an adapted and generalized version of the Heninger and Shacham key reconstruction algorithm~\cite{heninger2009reconstructing}.

The algorithm of Heninger and Shacham was already generalized by Bernstein et
al.~\cite{DBLP:conf/ches/BernsteinBGBHLV17}, but only to their special scenario,
in which an observation on the variables of the square-and-multiply algorithm was used.
Our approach is more generic and general. 
In the setting of cold-boot attacks, the generalizations by Henecka et al.~and Paterson et
al.~\cite{DBLP:conf/asiacrypt/PatersonPS12,DBLP:conf/crypto/HeneckaMM10} outperform the algorithm of Heninger and Shacham~\cite{heninger2009reconstructing}.
The main reason for this is that, given some partial information, there are some candidates compatible with this observation that are much more likely than other candidates. 
One can thus prune these unlikely candidates and only introduce a negligible error probability.
In contrast, in our scenario all of the candidates compatible with our observation are equally likely and no probabilistic pruning is possible.

\section{Conclusion}

We showed that side-channel resistance is not only relevant for cryptographic routines, but also for utility functions responsible for encoding and decoding secret data.
Nearly all of the major cryptographic libraries used lookup tables for these decoding purposes, allowing us to mount a high-resolution cache attack to significantly weaken the security guarantees provided by the underlying encryption schemes. 
We thus believe that it is important to check \emph{all} parts of a cryptographic library for side-channel vulnerabilities, e.\,g., by using automated analysis tools, especially for the case of strong attacker models enabled by trusted execution environments.

There are two important parameters making our attack feasible: 
First, the high resolution of our attack is possible only due to a security fix for transient execution attacks, as the serialization of memory loads greatly improves the signal-to-noise ratio.
Second, while the resulting leakage is quite small, the redundancy in the storage of the RSA keys allows us to achieve a significant drop in the security level of the secret key.
Both improvements, one which is intended to mitigate newly emerged attacks and the other targeting at speeding up RSA computations, come at the cost of security and in their combination render our attack possible.
We thus believe that studying performance optimizations, security patches and other improvements for their side effects is a crucial task and should

be conducted

continuously and across all functions which process sensitive data. We also propose to add side-channel analysis to the continuous integration pipelines, such that existing and newly introduced vulnerabilities are identified automatically, and known but minor leakages are re-evaluated depending on new developments in attack accuracy.

\section*{Acknowledgments}
The authors thank Nadia Heninger and Christopher Krebs for discussing approaches to integrate $q_{p}^{-1}$ into the key-recovery algorithm and the anonymous reviewers for their valuable comments and constructive feedback. This work has been supported by Deutsche Forschungsgemeinschaft (DFG) under grants 439797619 and 427774779.

\newpage

	\bibliographystyle{IEEEtran}
	{  
	  \footnotesize
	  \bibliography{IEEEabrv,main.bib}
	}

\newpage

\newpage
\begin{appendix}

  \section{Missing parts from Section~\ref{sec:key_recovery}}

  \subsubsection*{Blockwise Knowledge}
  We consider the situation that some blockwise knowledge about the secret key
  $\sk^{\star}=(p^{\star},q^{\star},d^{\star},d^{\star}_{p},d^{\star}_{q},{q^{\star}_{p}}^{-1})$ was obtained.
  In the following, we focus on the first five variables and treat $\sk^{\star}$ as a quintuple on the variables $\Vars=\{p,q,d,d_{p},d_{q}\}$.
  To simplify notation, for $v\in \Vars$, we denote the corresponding entry in some key $\sk$ by $\sk[v]$.
  We show in Sec.~\ref{app:last} that integrating the last variable $q_{p}^{-1}$ into the key-recovery approach does not directly give a usable linear equation in contrast to the other variables.

  In the situation given by our attack, we do not have observations on single bits, but on blocks consisting of $6$ bits, the length of a \basesf symbol.   
  In our model, we let $b\in \mathbb{Z}_{> 0}$ be the \emph{blocksize}. 
Without loss of generality, we assume that for each $v\in \Vars$, we have $b | \enc{\sk^{\star}[v]}$ by zero-padding all variables, i.\,e.~the length $\enc{\sk^{\star}[v]}$ of each variable $v$ in our secret key $\sk^{\star}$ is a multiple of~$b$. 
We denote the $i$-th bit of a bit-string $x$ by $x[i]$,
i.\,e.~the numerical value of $x$ is given by $\sum_{i=0}^{\enc{x}}2^{i}x[i]$.
The $j$-th \emph{block} $\block_{j}(x)\in \{0,\ldots,2^{b}-1\}$ of $x$ is defined as the value of the bitstring in the $j$-th block of $x$, i.\,e.~$\block_{j}(x)=\sum_{i=j\cdot b}^{j\cdot b+b-1}x[i]2^{i-j\cdot b}$. 
In our attack, we make use of the fact that the possible values for $\block_{j}(x)$ are partitioned into different sets to model the different cache lines used in our attack.
We consider a partition $\parts$ of the set $\{0,\ldots,2^{b}-1\}$, i.\,e.~$\parts$ is a set of sets $\parts_{1},\ldots,\parts_{|\parts|}$ such that
(i) $\bigcup_{i}\parts_{i}= \{0,\ldots,2^{b}-1\}$ and $\parts_{i}\cap
\parts_{i'}=\emptyset$ for all $i\neq i'$.
An \emph{observation} $\obs(\parts)$  with regard to
this partition $\parts$ is a quintuple that contains for each variable $v\in \Vars$ a vector in  $\{1,\ldots,|\parts|\}^{\enc{\sk^{\star}[v]}/b}$.
We denote the $j$-th entry of this vector by $\obs(\parts)[\sk^{\star}[v]]_{j}$. 
We say that an observation $\obs(\parts)$ is \emph{correct} for a secret key
$\sk^{\star}$ if for all $v\in \Vars$ and all $j\in
\{0,\ldots,(\enc{\sk^{\star}[v]}/b) - 1\}$, we have $\block_{j}(\sk^{\star}[v])\in \parts_{j'}$ with $j'= \obs(\parts)[\sk^{\star}[v]]_{j}$ if and only
if $\block_{j}(\sk^{\star}[v])\in \parts_{j}$. 

\subsection{Adapting the Algorithm}
The main idea of the algorithm is to reconstruct the different bits of the
secret key $\sk^{\star}$ iteratively.
We build up a set of \emph{candidates}.
Each candidate is a guess for the least significant bits of the true secret key $\sk^{\star}$ compatible with our observation and the RSA equations. 
We start our algorithm by producing a single candidate $\mywidetilde{\sk}$ of depth~$1$, i.\,e.~each variable only consists of a single bit.
We then apply the $\mathsf{expand}$ operation on $\mywidetilde{\sk}$ to obtain two candidates $\mywidetilde{\sk}_1$ and $\mywidetilde{\sk}_2$ of depth $2$ by using the RSA equations described by Heninger and Shacham~\cite{heninger2009reconstructing}.
Whenever a candidate has reached depth of a multiple of $b$, i.\,e., $j\cdot b$ for some $j$, we apply the $\mathsf{check}$ operation on this candidate to verify that the last produced block $\block_j(v)$ of each variable $v$ is feasible under our observation. 
If this candidate does not fit to our observation, we prune it.
We repeat these operations until a target depth~$D$ is reached.
All produced candidates of depth~$D$ are output. 
This target depth will be sufficient to reconstruct the remaining bits via the Coppersmith method~\cite{DBLP:journals/joc/Coppersmith97, DBLP:conf/ima/Howgrave-Graham97,DBLP:phd/de/May2003}. 
Informally, the depth of a candidate is the number of bits each variable has (see below for details).
Our algorithm performs these operations in a depth-first fashion (see 
Figure~\ref{fig:algorithm} in Sec.~\ref{sec:key_recovery}). 
We now give a more formal description of our algorithm. 
The $\mathsf{expand}$ operation uses a set of $4$ modular equations on $5$ variables and the $\mathsf{check}$ operation compares the generated candidates to our observations.
\begin{enumerate}[wide,  labelindent=0pt]
\item As a first step to set up our modular equations, we need to determine
  values $k$, $k_{p}$, and $k_{q}$ such that
  
    $e\cdot \sk^{\star}[d] = k(N-\sk^{\star}[p]-\sk^{\star}[q]+1)+1 $, 

    $e\cdot \sk^{\star}[d_{p}] = k_{p}(\sk^{\star}[p]-1)+1$, and
    $e\cdot \sk^{\star}[d_{q}] = k_{q}(\sk^{\star}[q]-1)+1$.
 
  We use the same technique as Heninger and
  Shacham~\cite{heninger2009reconstructing} to obtain these values.

\subsubsection*{Find $k$, $k_p$, and $k_q$}
An argument by Boneh, Durfee, and Frankel~\cite{DBLP:conf/asiacrypt/BonehDF98}
  shows that $0 < k < e$: As $d < (\sk^{\star}[p]-1)(\sk^{\star}[q]-1)=N-p-q+1$, having $k > e$ would be
  a contradiction to~$e\cdot \sk^{\star}[d] = k(N-\sk^{\star}[p]-\sk^{\star}[q]+1)+1 $.
  As $e=65537$ is by far the most common choice, we can thus enumerate all such values.
  For each such $k$, we can combine the three equations and easily compute the two solutions of the modular equation
  \begin{align*}
    x^{2} - [k(N-1)+1]\cdot x-k = 0 \pmod{e}. 
  \end{align*}
  This equation has two solutions $x_{1}$ and $x_{2}$ and it is easy to see that
  $\{x_{1},x_{2}\}=\{k_{p},k_{q}\}$ (see
  e.\,g.~\cite{heninger2009reconstructing}). 
  Hence, we can perform our algorithm on $(x_{1},x_{2})$ as well as on $(x_{2},x_{1})$ to determine
  the values $k$, $k_{p}$, and $k_{q}$ correctly. 
  
  With the above approach, we need to run the algorithm $2\cdot 65537$ times,
  which might take a long time.
  To rule out infeasible possibilities for $k$ earlier without running the complete algorithm, Boneh, Durfee, and
  Frankel~\cite{DBLP:conf/asiacrypt/BonehDF98} defined the value
  $\delta(\mywidetilde{k})= \lfloor(\mywidetilde{k}(N+1)+1) / e\rfloor$ for  $0 <
  \mywidetilde{k} < e$. 
  They then showed that for the correct value of $k$ corresponding to our secret key $\sk^{\star}$, we have $0\leq
  \delta(k)\leq \sk^{\star}[p]+\sk^{\star}[q]$.
  Hence, $\delta(k)$ and $\sk^{\star}[d]$ agree on the $\lfloor n/2 \rfloor -2$ most
  significant bits.
  We can thus compare, for each possibility $0 < \mywidetilde{k} < e$, the  most
  significant bits of $\delta(\mywidetilde{k})$ with the most significant bits given
  by our observations $\obs(\parts)[d]$ on $\sk^{\star}[d]$.
  If these do not agree, we discard our guess $\mywidetilde{k}$.
  Note that this reduces the running time significantly to only $2$ iterations of the algorithm, as almost always there
  is only one possible value left after this check. 

  In the following, we thus assume that we found the correct values for $k$,
  $k_{p}$, and $k_{q}$.
\item In order to iteratively add more bits to our candidates, we first need
  to find an initial candidate.
  As shown by Heninger and Shacham~\cite{heninger2009reconstructing}, we know that

    $e\cdot \sk^{\star}[d] \equiv 1 \pmod{2^{2+\tau(k)}}$, 
    $e\cdot \sk^{\star}[d_{p}] \equiv 1 \pmod{2^{1+\tau(k_{p})}}$, and 
    $e\cdot \sk^{\star}[d_{q}] \equiv 1 \pmod{2^{1+\tau(k_{q})}}$,
  
  where $\tau(x)$  is the exponent of the largest
  power of $2$ that divides $x$, i.\,e.~$\tau(x)=\max_{i}\{2^{i} | x\}$.
  Furthermore, both $\sk^{\star}[p]$ and $\sk^{\star}[q]$ are odd primes. 
  Hence, for our first candidate
  $\mywidetilde{\sk}$, we
  know the least significant bit of $\sk^{\star}[p]$, the least significant bit of $\sk^{\star}[q]$, the
  least significant $2+\tau(k)$ bits of $\sk^{\star}[d]$, the least significantly
  $1+\tau(k_{p})$ bits of $\sk^{\star}[d_{p}]$, and the least significantly $1+\tau(k_{q})$
  bits of $\sk^{\star}[d_{q}]$.

\item We say that a candidate $\mywidetilde{\sk}$ has \emph{depth} $i$, if the least
  significant $i$ bits of $p$ are set (and thus the least significant $i$ bits of $q$, the least significant $i+\tau(k)$ bits of $d$, the least significant $i+\tau(k_p)$ bits of $d_p$, and the least significant $i+\tau(k_q)$ bits of $d_q$). 
  Now, given a candidate
 $\mywidetilde{\sk}$ with
 depth $i$, we
 perform an $\mathsf{expand}$ operation, that produces two candidates of depth
 $i+1$.
 In order to do this, we need to determine the bits $p[i]$, $q[i]$,
 $d[i+\tau(k)]$, $d_{p}[i+\tau(k_{p})]$, and $d_{q}[i+\tau(k_{q})]$.
 Note that a trivial approach would continue the algorithm with all possible $2^{5}=32$
 assignments, but the partial knowledge given by our candidate allows us to
 drastically shrink the number of possibilities down to $2$.

  Therefore, we set up the following system of congruencies derived from the relations between the variables. 
  This is a system with $5$ variables and $4$ constraints and thus has exactly
  $2$ solutions.
  \begin{align*}
    p[i]+q[i]&\equiv \rhs_1[i] \pmod{2}\\
    d[i+\tau(k)]+p[i]+q[i] &\equiv \rhs_2[i+\tau(k)]  \pmod{2}\\
    d_{p}[i+\tau(k_{p})]+p[i] &\equiv \rhs_3[i+\tau(k_{p})] \pmod{2}\\
    d_{q}[i+\tau(k_{q})]+q[i] &\equiv \rhs_4[i+\tau(k_{q})] \pmod{2}
  \end{align*}
  Here, the right-hand sides are given as
  \begin{align*}
      \rhs_1 &= (N-\mywidetilde{\sk}[p]\cdot \mywidetilde{\sk}[q])\\
      \rhs_2 &= (k(N+1)+1-k(\mywidetilde{\sk}[p]+\mywidetilde{\sk}[q])-e\cdot \mywidetilde{\sk}[d])\\
      \rhs_3 &= (k_{p}(\mywidetilde{\sk}[p]+1)+1-e\cdot \mywidetilde{\sk}[d_{p}])\\
      \rhs_4 &= (k_{q}(\mywidetilde{\sk}[q]+1)+1-e\cdot \mywidetilde{\sk}[d_{q}]).
  \end{align*}
 
  Let $\mywidetilde{\sk}_{1}$ and $\mywidetilde{\sk}_{2}$ be the solutions of depth $i+1$
  obtained by 
  setting the position $i$ (resp.~$i+\tau(k)$,
  $i+\tau(k_{p})$, and$i+\tau(k_{q})$) of $\mywidetilde{\sk}$ to the solutions of the system. 
  For example, if $p[i]$ is part of the first solution, the candidate $\mywidetilde{\sk}_1[p]$
  for $p$ in $\mywidetilde{\sk}_{1}$ would be given by
  $\mywidetilde{\sk}_1[p]=\mywidetilde{\sk}[p]+2^{i}\cdot p[i]$. 

\item Now, whenever a candidate $\mywidetilde{\sk}$ of depth $j\cdot b+b-1$ is reached, we can
  check, whether the $j$-th block $\block_{j}(v)$ of each variable $v$ is feasible 
  under our observation $\obs$.
  We therefore check for each $v\in \Vars$, whether we have $\block_{j}(\mywidetilde{\sk}[v])\in \parts_{j'}$ with $j'= \obs(\parts)[\sk^{\star}[v]]_{j}$.  
  If this assignment is not possible, we prune the solution.
  We denote this check against our observation $\obs$ as $\mathsf{check}(\obs,\mywidetilde{\sk})$. 
\item Finally, whenever we find a candidate with our \emph{target depth} $D$, we
  output this candidate.
\end{enumerate}

We say that a candidate $\mywidetilde{\sk}$ of depth $i$ is \emph{compatible} with a secret key $\sk^{\star}$ if the $i$ (resp.~$i+\tau(k)$, $i+\tau(k_p)$, and $i+\tau(k_q)$) least significant bits of $\sk^{\star}[v]$ are identical to $\mywidetilde{\sk}[v]$ for all $v\in \Vars$. 
The correctness of the algorithm is easily seen by the following lemma.
\begin{lemma}
Let $\sk^{\star}$ be the correct secret key and $\mywidetilde{\sk}$ be a candidate of depth $i$ that is compatible with $\sk^{\star}$.
\begin{itemize}[wide,  labelindent=0pt]
    \item If $i=j\cdot b$ and $\obs$ is correct, $\mathsf{check}(\obs,\mywidetilde{\sk})$ will never prune $\mywidetilde{\sk}$.
    \item Let $\mywidetilde{\sk}_1$ and $\mywidetilde{\sk}_2$ be the output of $\mathsf{expand}(\mywidetilde{\sk})$.
    Then, either $\mywidetilde{\sk}_1$ or $\mywidetilde{\sk}_2$ are compatible with $\sk^{\star}$.
    \item The initial candidate of depth $1$ produced by the algorithm is compatible with $\sk^{\star}$.
\end{itemize}
\end{lemma}

\subsection*{Proof of Theorem \ref{thm:growth}}
\begin{proof}
  Expanding all of the candidates in $C$ with $b$ bits gives us exactly $2^{b}\cdot
  |C|$ incorrect candidates. 
  If we expand any incorrect candidate by $b$ bits, our assumption says that the
  blocks $j+1$ of these candidates behave like random $b$-bit strings.
  Fix one of these candidates $\mywidetilde{\sk}$.
  Now, $\mywidetilde{\sk}$
    is not pruned, if $\block_{j}(\mywidetilde{\sk}[v])\in \parts_{j'}$ with $j'= \obs(\parts)[\sk^{\star}[v]]_{j}$ for all $v\in \Vars$. 
    By our assumption, for each block, this happens with probability $\sum_{i=1}^{k}\left(
      |\parts_{i}|^{2}/2^{2b} \right)=2^{-H_{2}(\pr)}$, where
    $\pr[i]=|\parts_{i}|/2^{b}$.
    As these are independent, the probability that such an incorrect $\mywidetilde{\sk}$ is not pruned,
    is $2^{-5 H_{2}(\pr)}$.
    Hence, the expected number of non-pruned candidates where each block
    behaves like a random $b$-bit-string is exactly $|C|\cdot
    2^{b- 5\cdot H_2(\pr)}$.
    Furthermore, the expansion of the correct candidate gives us an additional
    $2^{b}-1$ incorrect candidates. 
\end{proof}

\subsection*{Proof of Theorem \ref{thm:final}}
\begin{proof}
  As noted above, we have $2^{b-1}\leq 2^{b}$ candidates of depth $b$.
  A simple induction  combined with Theorem~\ref{thm:growth}
  shows that the number of incorrect candidates with depth $j\cdot b$ is at most
  $ 2^b\cdot \sum_{i=0}^{j}(2^{b- 5\cdot H_2(\pr)})^{i} = 2^b\cdot \frac{(2^{b- 5\cdot H_2(\pr)})^{j+1}-1}{2^{b- 5\cdot H_2(\pr)}-1} $.
\end{proof}

\section*{The last parameter $q_{p}^{-1}$}
\label{app:last}
The attentive reader might have noticed that we obtain information about six parts of the secret key $p$, $q$, $d$, $d_p$, $d_q$, and $q_{p}^{-1}$, but do not use the information about $q_{p}^{-1}$ in our key reconstruction algorithm. 

In the following, we will shorty illustrate the problems of integrating $q_{p}^{-1}$ into the key-reconstruction algorithm.
First, note that, similar to the other variables of the secret key, one can easily conclude that there is some value $k'$ such that $q\cdot q_{p}^{-1} = k'\cdot p +1$. 

But the following adaption of an argument of Nguyen (described in~\cite{heninger2009reconstructing}) shows that knowing $k'$ already reveals the factorization of $N$.
As $q\cdot q_{p}^{-1} = k' \cdot p +1$, multiplying both sides of the equation by $p$ gives the equation $N\cdot q_{p}^{-1} = k'\cdot p^{2}+p$. 
Defining the polynomial $f(x)=k'\cdot x^2 +x$ shows that $f(p) \bmod N = 0$.
Hence, $p$ is a small root of a known polynomial (if $k'$ is known) and can thus be found by the method of Coppersmith~\cite{DBLP:journals/joc/Coppersmith97, DBLP:conf/ima/Howgrave-Graham97,DBLP:phd/de/May2003}.
\section{An example key in DER encoding}
\begin{figure}[h]

\begin{mdframed}[backgroundcolor=verylightgray]
\begin{lstlisting}[basicstyle=\scriptsize\ttfamily]
30 82 02 77 # SEQUENCE: Length 0x277
   02 01 00 # INTEGER: Version 00
   30 0d    # SEQUENCE: Length 0xd
      06 09 2a 86 ... 01 01 # Algorithm ID
      05 00
   04 82 02 61 # OCTET STRING: RSA Priv. Key
      30 82 02 5d # SEQUENCE: Length 0x25d
    	          # Private Key Parameters
         02 01 00              # Version 00
         02 81 81 00 ... a9 33 # n
         02 03 01 00 01        # e
         02 81 80 76 ... 79 a1 # d
         02 41 00 f3 ... e8 1f # p
         02 41 00 cf ... ac 6d # q
         02 40 2b 96 ... ef 8d # d mod (p-1)
         02 41 00 c0 ... 85 95 # d mod (q-1)
         02 41 00 89 ... 8c 19 # q^-1 mod p
\end{lstlisting}
\end{mdframed}
\caption{1024-bit RSA private key, DER encoded according to PKCS \#8, in hexadecimal format.}
\label{lst:rsa_key_der}
\end{figure}

\newpage
\section{Leakage estimation with Microwalk}
\begin{table}[h]
    \caption{The leakage estimation from \emph{Microwalk} for \openssl, generated from 4,096 test cases. For each instruction, \emph{Microwalk} computes the Mutual Information (MI) between the memory access traces and the test case IDs, which measures the ability of an attacker to infer the input from an observed trace. Note that the leakage is upper bounded by the logarithm of the number of test cases (12).}
    \label{tab:microwalk-results}
    \centering
    {\small
    \begin{tabular}{lll}
        Instruction & Avg. leakage (bits) \\ \hline 
        \texttt{EVP\_DecodeUpdate+105} & 12 \\
        \texttt{EVP\_DecodeBlock+E} & 12 \\
        \texttt{EVP\_DecodeBlock+59} & 12 \\
        \texttt{EVP\_DecodeBlock+C0} & 12 \\
        \texttt{EVP\_DecodeBlock+D4} & 12 \\
        \texttt{EVP\_DecodeBlock+E5} & 12 \\
        \texttt{EVP\_DecodeBlock+FA} & 12 \\
        \texttt{BN\_bin2bn+AC} & 4.035 \\
        \texttt{BN\_bin2bn+24} & 4.016 \\
        \texttt{ASN1\_get\_object+171} & 2.965 \\
        \texttt{ASN1\_get\_object+16} & 2.941 \\
        \texttt{ASN1\_get\_object+CC} & 2.941 \\
        \texttt{PEM\_read\_bio+24E} & 1.012 \\
        \texttt{EVP\_DecodeUpdate+E2} & 1.012 \\
        \texttt{EVP\_DecodeUpdate+F8} & 1.012 \\
        \texttt{EVP\_DecodeBlock+49} & 1.012 \\
        \texttt{EVP\_DecodeBlock+14D} & 1.012 \\
        \texttt{EVP\_DecodeBlock+C8} & 1.012 \\
        \texttt{EVP\_DecodeBlock+D9} & 1.012 \\
        \texttt{EVP\_DecodeBlock+EE} & 1.012 \\
        \texttt{EVP\_DecodeBlock+138} & 1.012 \\
        \texttt{EVP\_DecodeBlock+142} & 1.012 \\
        \texttt{EVP\_DecodeBlock+148} & 1.012 \\
        \texttt{PEM\_read\_bio+1D0} & 1.009 \\
        \texttt{PEM\_read\_bio+1ED} & 1.009 \\
        \texttt{PEM\_read\_bio+1F2} & 1.009 \\
    \end{tabular}}
\end{table}

\newpage
\section{BoringSSL's \basesf decoding}
\begin{figure*}[b]
\begin{mdframed}[backgroundcolor=verylightgray]
\begin{lstlisting}[basicstyle=\scriptsize\ttfamily,language=C++]
static uint8_t base64_ascii_to_bin(uint8_t a) {
  // Since PEM is sometimes used to carry private keys, we decode base64 data
  // itself in constant-time.
  const uint8_t is_upper = constant_time_in_range_8(a, 'A', 'Z');
  const uint8_t is_lower = constant_time_in_range_8(a, 'a', 'z');
  const uint8_t is_digit = constant_time_in_range_8(a, '0', '9');
  const uint8_t is_plus = constant_time_eq_8(a, '+');
  const uint8_t is_slash = constant_time_eq_8(a, '/');
  const uint8_t is_equals = constant_time_eq_8(a, '=');
  uint8_t ret = 0xff;  // 0xff signals invalid.
  ret = constant_time_select_8(is_upper, a - 'A', ret);       // [0,26)
  ret = constant_time_select_8(is_lower, a - 'a' + 26, ret);  // [26,52)
  ret = constant_time_select_8(is_digit, a - '0' + 52, ret);  // [52,62)
  ret = constant_time_select_8(is_plus, 62, ret);
  ret = constant_time_select_8(is_slash, 63, ret);
  // Padding maps to zero, to be further handled by the caller.
  ret = constant_time_select_8(is_equals, 0, ret);
  return ret;
}
\end{lstlisting}
\end{mdframed}
\caption{Base64 decoding constant-time implementation in Google's \boringssl \cite{googleboringssl} (crypto/base64/base64.c)} \label{lst:constant_time_boringssl}
\vspace{14cm}
\end{figure*}

\newpage

\end{appendix}

\end{document}